\documentclass{aa}  

\usepackage{graphicx}
\usepackage{txfonts}
\usepackage{hyperref}
\hypersetup{
  colorlinks   = true, %Colours links instead of ugly boxes
  urlcolor     = blue, %Colour for external hyperlinks
  linkcolor    = blue, %Colour of internal links
  citecolor   = blue %Colour of citations
}
\usepackage{orcidlink}
\usepackage{xcolor}

\newcommand{\commentzk}[1]{{\color{violet} (#1)}}

\newcommand\romanst{\textit{Roman}}

\newcommand{\ari}{Zentrum f{\"u}r Astronomie der Universit{\"a}t Heidelberg, Astronomisches Rechen-Institut, M{\"o}nchhofstr. 12-14, 69120 Heidelberg, \\ Germany}
\newcommand{\llnl}{Space Science Institute, Lawrence Livermore National Laboratory, 7000 East Ave., Livermore, CA 94550, USA}

\begin{document}

    \title{Astrometric microlensing by isolated black holes in the Roman Galactic Bulge Time Domain Survey}
    \titlerunning{Astrometric microlensing by isolated black holes with Roman}
    \author{Zofia Kaczmarek\inst{1}
    \and
    Peter McGill\inst{2}
    }

    \institute{
    \ari
    \and
    \llnl
    }

       \date{Received XXX; accepted YYY}
       
\abstract{Most of the hundred million stellar-mass black holes (BHs) expected in the Milky Way are predicted to be isolated. Characterizing this BH population will provide insights into stellar evolution and supernova explosion physics. The Nancy Grace Roman Space Telescope Galactic Bulge Time Domain Survey (GBTDS) will be an unprecedented astrometric survey. It will enable detection of isolated BHs through astrometric microlensing, apparent stellar position shifts caused by deflection of distant starlight.}
{
We assess the GBTDS's expected yield, sensitivity, characterizing power, and signal characteristics of isolated BHs detectable through astrometric microlensing. This work aims to inform strategies to find these events in the GBTDS data. 
}
{
We combine a Galactic model, stellar population synthesis model, BH initial-final mass relation, and kick velocity distribution to simulate lensing isolated BHs. We simulate GBTDS astrometry to assess which BHs are detectable and can be characterized.
}
{
We predict $ 487 \pm 37 $ isolated-BH microlensing events with detectable astrometric signals, 82\% of which have impact parameters beyond 2 Einstein radii and consequently have small or undetectable photometric signals. Standard astrometric fits yield unit weight errors $ >1.2 $ for 18\% and $ >2 $ for 2\% of these events. Sensitivity peaks for $ 5 $--$ 6\,M_\odot $ BHs and is limited by the five-year GBTDS baseline. Weak microlensing parallax constraints limit BH mass precision, but events with Einstein timescales $ >100\,\mathrm{days} $ and angular Einstein radii $ >1.2\,\mathrm{mas} $ have high BH lens probabilities within our Galactic model.
}
{
Wide-separation astrometric microlensing offers an independent discovery channel for isolated BHs beyond photometrically selected events. Realizing this opportunity requires dedicated multi-season GBTDS searches, joint astrometric and photometric modeling, and accounting for other astrometric variability. Although precise individual masses are challenging for these events, probabilistic lens classification could lead to Galactic abundance constraints on the population of isolated BHs.
}

\maketitle

\section{Introduction} \label{sec:intro}

The Milky Way is expected to harbour about one hundred million black holes (BHs) that formed during the death and collapse of massive stars \citep[e.g.,][]{Samland1998, AgKa2002, Elbert2018, Olejak2020}. Characterizing this population will provide insights into stellar evolution mechanisms \citep[e.g.,][]{Miller-Jones2021, Ivanova2013, Rose2022}, supernova explosion physics \citep[e.g.,][]{Sukhbold2016, Repetto2017}, and the means to understand the extragalactic population of massive merging binary BHs detected via gravitational-wave emission \citep[e.g.,][]{Abbott2023, Ho2024}.

Despite their expected abundance, little is known about the Galactic BH population. This is due in large part to the difficulty of detecting these dark objects. All but one of the known $\sim50$ Galactic BHs have been detected via interactions with a companion in binary systems, e.g., astrometric \citep{Miller-Jones2021, El-Badry2023, El-Badry2023a, Panuzzo2024} or X-ray \citep{Remillard2006,Corral2016} binaries. These detection methods are sensitive only to specific regions of parameter space, e.g., BH and companion masses and orbital parameters, leading to strong selection effects.

This limited sample of BHs leaves a number of open questions about their demographics. A notable example is the ``mass gap'' observed between known neutron stars (NSs) and BHs \citep{Bailyn1998, Ozel2010, Farr2011}. It remains unclear whether the gap is caused by astrophysical effects preventing the formation of remnants within this mass range \citep[e.g.,][]{Belczynski2012, Fryer2012, Olejak2022} or by possible observational biases \citep[e.g.,][]{Kreidberg2012, WyrzykowskiMandel2020, Kruszynska2024}. Furthermore, in recent years, a new, numerous population of BHs in the mass range of 30--150 $M_\odot$ has been discovered via gravitational-wave observations \citep[][]{Abbott2016, Abbott2017, Abbott2021b, Abbott2023, Abbott2021, LIGO2025, Abac2025}. These results need to be reconciled with the much lower BH masses observed in our Galaxy \citep[e.g.,][]{Corral2016} as well as those predicted by stellar evolution models \citep[e.g.,][]{Sukhbold2016}. The recent discovery of a 33 $M_\odot$ BH in the Milky Way \citep{Panuzzo2024} has renewed interest in consistent models that could explain the properties of BHs discovered through different observational channels. Importantly, the vast majority of stellar-origin BHs in the Galaxy are expected to be isolated \citep{Lamberts2018, Olejak2020}, meaning that the sample of known BHs in binary systems is unlikely to be representative of the Galactic population. Therefore, to build a complete and unbiased census of BHs in the Galaxy, the sample of isolated BHs must be increased.

The only currently known isolated stellar-mass black hole was detected recently via a gravitational microlensing event, OB110462 \citep{Sahu2022, Lam2022, Mroz2022, LamandLu2023}. During this event, the black hole gravitationally deflected the light of a distant background source, causing a source flux amplification \citep[photometric microlensing;][]{Paczynski1986largeopticaldepth} and positional shift \citep[astrometric microlensing;][]{Walker1995, Miyamoto1995, Hog1995}. Using both of these signals, the lens mass was determined to be approximately $7 M_{\odot}$. This event was initially reported by the Optical Gravitational Lensing Experiment \citep[OGLE;][]{Udalski2015}. Indeed, there are many examples of isolated black hole candidates that have been found via photometric microlensing events \citep[e.g.,][]{Wyrzykowski2016, WyrzykowskiMandel2020, Kruszynska2024, Rybicki2024, Howil2025, Kaczmarek2025}. However, such candidates cannot be definitively confirmed unless further follow-up, astrometric or otherwise, is carried out.

Astrometric microlensing allows a measurement of the angular Einstein radius that is crucial for measuring the lens mass \citep{Hog1995, Walker1995, Miyamoto1995, Dominik2000} and is not possible with photometric data alone. Surveys combining astrometric and photometric microlensing observations could be especially powerful in providing precise mass measurements for dark remnants \citep{Rybicki2018, Kaczmarek2022, Sajadian2023}. Additionally, astrometric microlensing acts at larger lens--source separations than photometric microlensing \citep{Miralda-Escude1996}; astrometric microlensing also has longer and more mass-sensitive timescales \citep{Honma2001}. At the GBTDS's expected astrometric precision, for some classes of lenses, e.g., massive isolated BHs, there will be events with detectable astrometric signatures but small or even undetectable photometric signatures \citep[e.g.,][]{Fardeen2024}. Therefore, including astrometry-only detections would increase the sample sizes of microlensing events and could improve population studies, which is particularly valuable for populations with very few known objects, such as isolated stellar-mass BHs.

Observations of astrometric microlensing show great promise for characterizing stellar remnants \citep{Sahu2017, McGill2023}, but remain challenging due to the required precision. Even the most favourable events produce signals of $\sim$1 milliarcsecond (mas) and are detectable only with a few observatories. All detections to date have required multi-year astrometric follow-up at this precision, which is difficult to access and highly oversubscribed \citep[see e.g.,][]{Kaczmarek2025}. 
However, we are now entering an era of astrometric surveys providing regular time-series data precise enough to detect shifts caused by microlensing. For example, \textit{Gaia} \citep{Prusti2016} achieved 0.1--1\,mas single-observation precision \citep{Lindegren2021}, with tentative astrometric microlensing signals already reported \citep[e.g.,][]{Jablonska2022} and astrometry-only detection and modelling methods proposed for future releases \citep{Chen2023, Jankovic2025}. However, crowding limits Gaia's performance in fields where most microlensing events occur \citep[e.g.,][]{Luna2023, CastroGinard2024}.

Launched in August 2026, the \textit{Roman Space Telescope} \citep[hereafter \romanst]{Spergel2015} is the first space mission with microlensing as one of its key science goals. \romanst~will monitor the Galactic Bulge with the Wide Field Instrument (WFI) over 5 years of its primary mission performing the Galactic Bulge Time Domain Survey (GBTDS). The survey design is optimized for microlensing science, focusing primarily on cold exoplanet demographics \citep[][]{ROTAC}. GBTDS targets a $\approx 1.68$ deg$^2$ field near the Galactic Center observing in a wide near-infrared filter ($F146$) that minimizes the effects of extinction, while also achieving high photometric precision by maximising collected flux \citep{Penny2019, McKinnon2026}. Exposures will be obtained every 12.1 minutes with a photometric precision of $\sim$10 mmag, in a design tailored for large-scale microlensing exoplanet detection \citep{Penny2019, Wilson2023}. Due to WFI's expected stability and precise tracking, each of these high-cadence exposures will also provide position measurements at a precision of about 1 mas; exposures could be stacked for a further significant improvement in precision \citep[e.g.,][]{McKinnon2026, Lam2026}. However, microlensing exoplanets produce small astrometric signals with angular scales of microarcseconds \citep{Safizadeh1999}; consequently, anomalous astrometry will not be processed in \romanst's microlensing pipeline \citep{RomanMSOSForROSES}. That means additional efforts must be made to identify astrometric microlensing events in the GBTDS data. 

In this study, we simulate and analyse GBTDS observations of the population of isolated BHs that can be detected through astrometric microlensing. We extend previous work on detecting microlensing isolated BHs that considers only small lens--source impact parameter events (typically < 2 angular Einstein radii) with both detectable astrometric and photometric signals \citep[][]{Pruett2022, Sajadian2023} to the wider lens--source impact parameter regime \citep[e.g.,][]{Tilburg2018, Verma2023, Fardeen2024, Jankovic2025}.

The paper is structured as follows. In Section~\ref{sec:astrometric_microlensing}, we cover the astrometric microlensing effect. In Section~\ref{sec:simulation}, we outline the process of simulating the GBTDS and detecting astrometric microlensing events. In Section~\ref{sec:results}, we report yields of detectable astrometric microlensing events caused by isolated BH lenses and we analyse the constraints that can be placed on physical properties of the lenses. We also investigate characteristics of the astrometric microlensing signals and how they can be selected from the GBTDS. Finally, in Section~\ref{sec:discussion}, we discuss and summarize our findings, including recommendations for finding isolated BHs in GBTDS.

\section{Astrometric microlensing}
\label{sec:astrometric_microlensing}

Consider a lens of mass $M_{L}$ at distance $D_{L}$ and a more distant background source at distance $D_{S}$ from an observer. If the source, the lens, and the observer are perfectly collinear, an Einstein ring image of the source is formed with angular radius $\theta_\mathrm{E}$ \citep{Einstein1936};
% ^v
\begin{equation}
\theta_\mathrm{E} = \sqrt{\frac{4GM_{L}}{c^2}\frac{D_\mathrm{S} - D_\mathrm{L}}{D_\mathrm{S}D_\mathrm{L}}}\,.
\label{eq:theta_E}
\end{equation}
If the lens--source alignment is imperfect, i.e., the lens and source have an angular separation $\boldsymbol{u}$, pointing towards the source position from the lens position and normalized by $\theta_{\rm E}$, two images of the source are formed. If the images are unresolved and the lens is dark, the apparent astrometric shift of the source-image centroid from the unlensed source position \citep[see e.g.,][]{Bramich2018} is:
\begin{equation}
\boldsymbol{\delta}(\boldsymbol{u}) = \frac{\boldsymbol{u}}{|\boldsymbol{u}|^{2}+2}\theta_{\rm E}.
\label{eq:full_shift}
\end{equation}
The apparent photometric amplification of the source flux is \citep[e.g,][]{Paczynski1986largeopticaldepth},
\begin{equation}
    A = \frac{|\boldsymbol{u}|^2 + 2}{|\boldsymbol{u}|\sqrt{|\boldsymbol{u}|^2 + 4}}.
    \label{eq:ampl}
\end{equation}
As the lens--source separation changes over the course of an event, it can be parameterised as:
\begin{equation}
    \boldsymbol{u}(t) = \boldsymbol{u_{0}} + \frac{(t - t_{0})}{t_{\rm E}}\boldsymbol{\hat{\mu}_{\text{rel}}} + \boldsymbol{P}(t; \boldsymbol{\pi_{\rm E}})\,.
    \label{eq:lens_source_track}
\end{equation}
Here, $\boldsymbol{u_{0}}$ is the angular lens--source impact parameter, pointing towards the source position and normalized by $\theta_{\rm E}$; hereafter, $u_{0}=|\boldsymbol{u_{0}}|$. $t_{0}$ is the moment of closest approach, $\boldsymbol{u}(t_{0}) = \boldsymbol{u_{0}}$, $\boldsymbol{\hat{\mu}_{\text{rel}}}$ is the unit vector in the direction of the relative lens--source proper motion, and $\boldsymbol{P}$ is the relative lens--source parallax motion.

The parameters that contain information on the lens and source are the Einstein timescale and microlensing parallax, defined as
\begin{equation}
    t_{\rm E} = \frac{\theta_{\rm E}}{\mu_{\text{rel}}},\quad{}\quad{}|\boldsymbol{\pi_{\rm E}}|=\frac{\rm AU}{\theta_{\rm E}}\left(\frac{1}{D_{L}}-\frac{1}{D_{S}}\right),
    \label{eq:tE_piE}
\end{equation}
respectively. Here, $\mu_{\rm rel}=|\boldsymbol{\mu_{\rm rel}}|$, and AU is one astronomical unit. If $\theta_{\rm E}$ and $\pi_{\rm E}$ can be constrained, the lens mass can be measured \citep[e.g.,][]{Gould2000},
\begin{equation}
    M_{L}=\frac{\theta_{\rm E}}{\kappa\pi_{\rm E}}, \quad \kappa\approx 8.14\text{ mas}/M_{\odot}.
\end{equation}
For $u\gg\sqrt{2}$, assuming a detectability threshold $\delta_T$, the region around the lens where a background source would cause a detectable signal can be approximated by a circle of radius $u_{T} =\theta_{\rm E}/\delta_{T}$ \citep[e.g.,][]{Dominik2000, Bramich2018}. Assuming a one-day stack of GBTDS exposures corresponding to an astrometric precision of $\sim$0.1 mas for a $\sim$21 F146-band magnitude source \citep[e.g.,][]{Lam2026}, and $\theta_{\rm E}\approx1$ mas for a BH, this gives $u_{T}\sim10$, which is greater than the lens--source separation regimes considered previously \citep[e.g.,][]{Sajadian2023}. The timescale of the astrometric microlensing effect can be quantified as the time spent by the source, neglecting parallax motion, within an angular radius of $u_{T}$ of the lens \citep{Honma2001},
\begin{equation}
t_{\text{ast}} = 2t_{\rm E}\sqrt{u^{2}_{T}-{u_{0}}^{2}}\,.
\label{eq:tast} 
\end{equation}
For a given survey duration $T_{\rm obs}$ and assuming sufficiently high cadence, if an event peaks within $T_{\rm obs}$ and $t_{\rm ast}<T_{\rm obs}$, the survey will observe its full amplitude from peak astrometric deviation to baseline. However, for typical lenses in the Galaxy, $t_{\rm ast}$ can be $\sim1$--$10$ years \citep[e.g.,][]{Belokurov2002} and exceed the survey duration, which means that some long-duration events will not be detected solely because of an insufficient temporal baseline. Specifically, we must require some threshold change in the astrometric signal as a detection \citep[e.g.,][]{Dominik2000}. This point is relevant for the 5-year duration of the GBTDS, and we therefore consider it in our event selection criteria. It is also worth mentioning that blended light from unrelated neighbouring sources changes the form of the centroid shift in Eq. (\ref{eq:full_shift}) and also reduces the magnitude of the astrometric microlensing signal \citep{Bramich2018,Lam2020}. To avoid this complication, we limit our sample of events to those with no or low blending. 

\iffalse
For events with $t_{\rm ast}>T_{\rm obs}$, the area around the lens where a background source would have a change in astrometric \commentzk{?} within $T_{\rm obs}$ can be approximated by a circle of radius \citep{Dominik2000, Fardeen2024},
\begin{equation}
    u_{\Delta} \approx \sqrt{\frac{T_{\text{obs}}\theta_{\rm E}}{\delta_{T}t_{\rm E}}}\,.
    \label{eq:u_delta}
\end{equation}
%
\commentzk{we might need to rewrite this part - I don't know if we're using $t_{ast}$ for selection?}
For events with a luminous lens, stars and white dwarfs, flux from the lens acts to dilute the astrometric signal and decrease the detectability to circles of radii $u^{\rm lum}_{T} = u_{T}/(1+g)$ and $u^{\rm lum}_{\Delta} = u_{\Delta}/\sqrt{1+g}$ for a threshold signal amplitude and change in signal threshold with $T_{\rm obs}$, respectively \citep{Dominik2000}. Blending light from unrelated neighbouring sources changes the form of the centroid shift in Eq. (\ref{eq:full_shift}) and also acts to reduce the magnitude of the astrometric microlensing signal \citep{Bramich2018,Lam2020}.
\fi

\section{Simulation}
\label{sec:simulation}

\subsection{Galactic model, stellar populations, and remnants}

To simulate a catalogue of microlensing events, we used the Population Synthesis for Compact Lensing Events code \citep[\texttt{PopSyCLE};][]{Lam2020}. 
\texttt{PopSyCLE} uses a modified version of \texttt{Galaxia} (see \citealt{Sharma2011}, Appendix of \citealt{Lam2020}) for the underlying stellar model of the Milky Way that is based on the Besançon model \citep{Robin2004}. 
BHs, NSs and WDs are injected into the simulation by evolving clusters matching each \texttt{Galaxia} stellar sub-population, assuming they are single-age, single-metallicity populations, via the Stellar Population Interface for Stellar Evolution and Atmospheres code \citep[\texttt{SPISEA};][]{Hosek2020}. 
BHs and NSs are given natal kick velocities; initial--final mass relations (IFMRs) are used to determine masses of BHs, NSs and White dwarfs (WDs) given the physical properties of their progenitors. 
The Galactic model parameters and relations adopted for the simulation are detailed in Table \ref{tab:simulation}.

\begin{table}[t!]
\begin{center}
\caption{\label{tab:simulation} Summary of \texttt{PopSyCLE} simulation parameters.}
\begin{tabular}{p{40mm}p{38mm}}
    \hline\hline 
    Parameter & Value \\\hline
    Milky Way escape velocity & $550$kms$^{-1}$ \citep{Piffl2014} \\
    Sun--Galactic Center distance & $8.3$kpc \\
    Mean of initial BH progenitor kick distribution & $100$kms$^{-1}$ \\
    Mean of initial NS progenitor kick distribution & $350$kms$^{-1}$ \\
    Initial--Final Mass Relation & Sukhbold N20 \citep[based on models from][]{Sukhbold2014, Sukhbold2016, Woosley2017, Woosley2020} \\
    Bar dimensions (radius, major axis, minor axis, height) & $(2.54, 0.70, 0.424, 0.424)$ kpc \\
    Bar angle (Sun--Galactic Center, 2nd, 3rd) & $(62.0, 3.5, 91.3)$ $^{\circ}$ \\
    Bulge velocity dispersion (radial, azimuthal, z) & $(100, 100, 100)$ kms$^{-1}$ \\
    Bar pattern speed & 40.00kms$^{-1}$ kpc$^{-1}$ \\
    \hline
\end{tabular}
\end{center}
\tablefoot{The bulge parameters were chosen to be consistent with the ``v3'' version in \citet[App.~A]{Lam2020}, and match the event rates reported by OGLE~\citep{Mroz2019}.}
\end{table}

\subsection{Galactic Bulge Time Domain Survey}\label{sec:sim}

\begin{figure}[t!]
    \includegraphics[width=\columnwidth, trim=0cm 0cm 0cm 0cm]{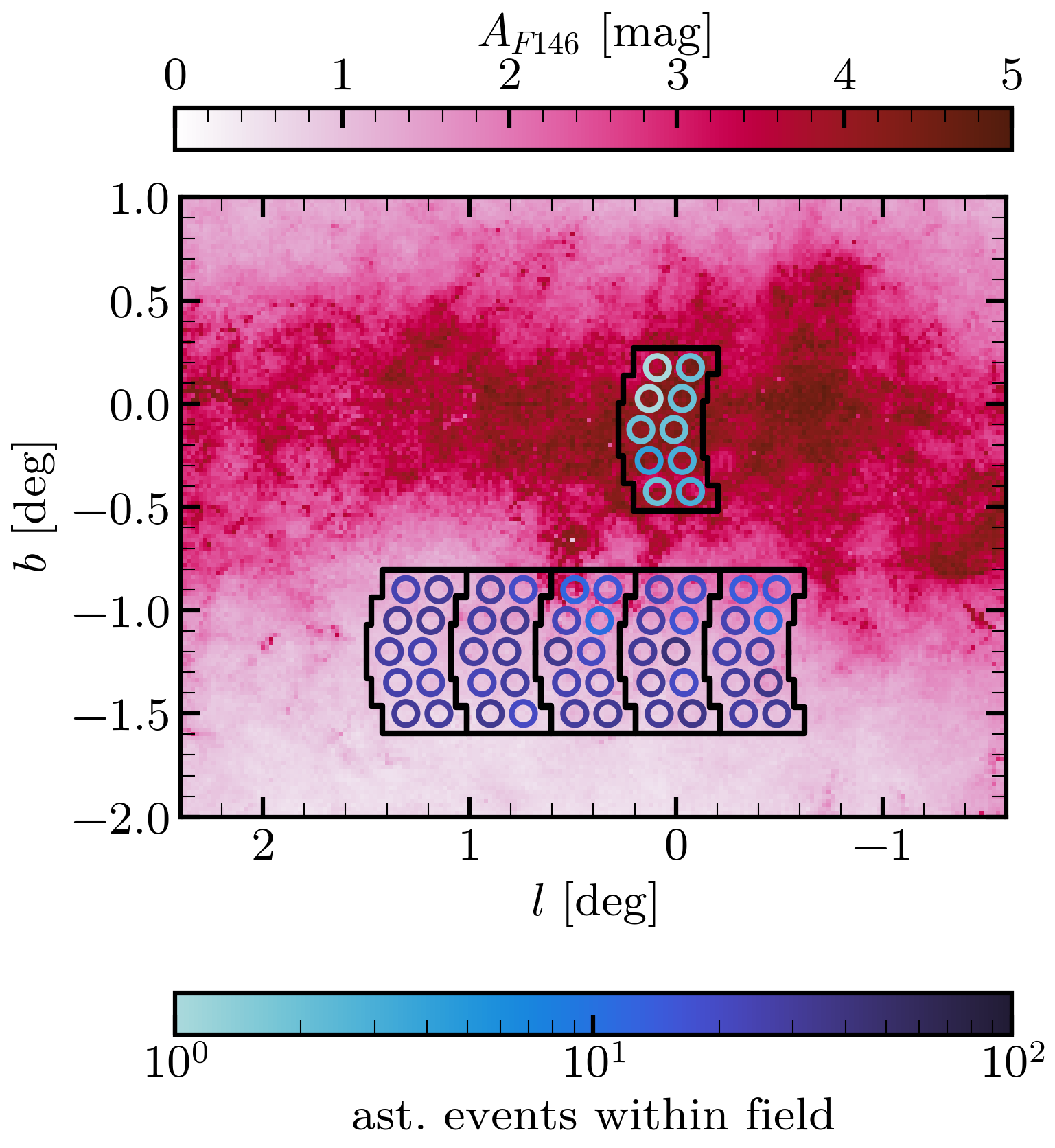}
    \caption{Layout of the simulation fields (blue circles), overplotted on the $F146$ extinction map \citep[magenta;][implementation discussed in Section~\ref{sec:sim}]{Surot2020, Huston2026}. Shades of blue of the field outlines represent the total number of detectable astrometric microlensing events within each simulation field, while shades of magenta in the background pixel map represent extinction. The event yields are heavily impacted by extinction, as lowest-yield fields overlap with high-$A_{F146}$ regions. The nominal \romanst~GBTDS footprint has been overplotted with black solid lines using the {\tt gbtds\_optimizer} software\textsuperscript{a}.}
    \tiny{\textsuperscript{a} 
    \url{https://github.com/mtpenny/gbtds_optimizer}}
    \label{fig:footprint}
\end{figure}

Our simulated survey largely follows the GBTDS simulation of \citet{Kaczmarek2026}. The main difference introduced in this work is a much wider permitted range for the impact parameter $u_0 \in (-10, 10)$, allowing for capturing events with no photometric signal detectable only in astrometric time series. We approximate the GBTDS footprint by simulating ten separate fields of $0.01$ deg$^{2}$ within each of the six WFI fields (see Fig.~\ref{fig:footprint}). Lensing events are selected from the simulated populations as pairs of luminous sources and foreground lenses that reach a separation of $\leq$ 10$\theta_{\rm E}$ within the simulated survey time of 5 years (\romanst 's primary mission\footnote{\url{https://roman-docs.stsci.edu/roman-community-defined-surveys/galactic-bulge-time-domain-survey}}). In all the work that follows, predicted event rates are scaled to the full $1.68$ deg$^{2}$ GBTDS area as defined in the Roman Observations Time Allocation Committee Final Report and Recommendations \citep[][hereafter: ROTAC report]{ROTAC}.

Lens and source magnitudes reported by \texttt{PopSyCLE} in the UBV, $H$, $J$, and $K$ filters are converted to extinction-free $F146$-band magnitudes using the relations of \citet{Wilson2023}. Extinction models performing badly near the Galactic Center are a known problem \citep[e.g.,][]{Gonzalez2012, Huston2026}. Therefore, we build on \texttt{SynthPop}\footnote{\url{https://github.com/synthpop-galaxy/synthpop/}} \citep{Kluter2025, Huston2026} and its implementation of the \citet{Surot2020} extinction map. \citet{Surot2020} have created a two-dimensional high-resolution $E(J - K_s)$ map based on VISTA Variables in the Via Lactea (VVV) observations of red clump and red giant branch stars. \citet{Huston2026} have implemented this map as a query-able ${A_K}_s(l,b,d)$ function by scaling the 2D extinction with the line-of-sight dust distribution; they specifically choose the \citet{Surot2020} map for reproducing \textit{Roman}'s GBTDS fields, optimizing the model for bulge stars. As implemented in {\tt SynthPop}, we take:
\begin{equation}
    A_\lambda/{A_K}_s=(\lambda/\lambda_{K_s})^{-\alpha}
\end{equation}
where $\alpha = 2.255$, $\lambda_{K_s} = 2.152$~{\textmu}m. As $F146$ is a very wide filter, we integrate over its extents. For simplicity, we assume the $F146$ passband to be a tophat with transmission of 1 for $\lambda \in (0.927, 2.0)$ {\textmu}m and 0 otherwise. We assume the incoming spectrum to be that of a blackbody with $\log_{10} (T_{\rm eff}) =  3.74$, the median of our simulated source stars (with a 10$^\text{th}$-90$^\text{th}$ percentile range of $3.65 - 3.80$; changing $T_{\rm eff}$ within those ranges only introduces a $\lesssim$ 0.1 mag error in the calculated $A_{F146}$ values). With photon-counting weights,
\begin{equation}
    A_{F146}({A_K}_s) = -2.5\log_{10}
    \left(  \frac{\int B_\lambda(T_{\rm eff}) \lambda 10^{-0.4 {A_K}_s(\lambda/\lambda_{K_s})^{-\alpha}} d\lambda}{\int B_\lambda(T_{\rm eff}) \lambda \ d\lambda}\right).
\end{equation}

We realise this integration on a 2000-point grid uniform in $\lambda$, and map ${A_K}_s(l,b,d)$ to $A_{F146}(l, b, d)$. We show the resulting $A_{F146}(l, b, d)$ map for a fixed $d = 8.15$ kpc -- the Galactic center distance assumed in \texttt{SynthPop} -- in Fig.~\ref{fig:footprint}.

We adopt the nominal GBTDS cadence pattern detailed in the ROTAC report, which follows Galactic bulge visibility seasons centered around the vernal and autumnal equinoxes. Six high-cadence observing seasons are defined, each lasting 70.5 days; within each of the seasons, observations are taken at a $12.1$ minute cadence in the $F146$ bandpass over the full survey area. Seasons three and four are spaced with a `large season gap' of 4 Galactic bulge visibility seasons. We denote this set of $50\,346$ observing times over the GBTDS $\mathcal{T}_{\text{GBTDS}}$.

\begin{table*}[]
\begin{center}
\caption{Prior distributions for modeling the detected events.}
\begin{tabular}{p{1.3 cm} p{7.5 cm} p{1.8 cm} p{6cm}}
\hline \hline
Parameter & Prior distribution & Unit & Definition \\
\hline
$u_0$ & $\mathcal{U}(-10, 10)$ & -- & impact parameter\\
$t_0$ & $\mathcal{U}(\min(\mathcal{T}_{GBTDS}), \max(\mathcal{T}_{GBTDS}))$ & decimal year & time of closest approach (rel. lin. motion)\\

$\log_{10}(t_{\rm E})$ & $\mathcal{U}(-2, 0.5)$ & years & log (event timescale)\\
$\log_{10}(\pi_{\rm E})$ &  $\mathcal{U}(-2.5, 0)$ & -- & log (microlensing parallax)\\
$\log_{10}(\theta_{\rm E})$ & $\mathcal{U}(-1, 1)$ & mas & log (Einstein radius)\\

$\mu_{\alpha*,S}$ & $\mathcal{N}(-3, 5)$ truncated to (-18, 12)& mas/yr & source proper motion in $\alpha \cos \delta$\\
$\mu_{\delta,S}$ & $\mathcal{N}(-5, 5)$ truncated to (-20, 10) & mas/yr & source proper motion in $\delta$\\

$\log_{10}\pi_S$ & $\mathcal{U}(-2, 0)$ & mas & log (source parallax)\\
$\Delta \alpha_{0,S}$ & $\mathcal{N}(\text{mean, std})$ truncated to \text{(min, max)} of $\Delta \alpha^*$ data & mas & offset to source ref. position in $\alpha \cos \delta$\\
$\Delta \delta_{0,S}$ &  $\mathcal{N}(\text{mean, std})$ truncated to \text{(min, max)} of $\Delta \delta$ data  & mas & offset to source ref. position in $\delta$\\
\hline
$F_{146,0}$ & $\mathcal{N}(\text{median, std})$ truncated to \text{(min, max)} of $F_{146}$ data & mag & baseline magnitude (optional) \\
\hline
\label{tab:constraints_priors}
\end{tabular}
\tablefoot{$\mathcal{U}(a,b)$ is the uniform distribution with lower and upper bounds of a and b, respectively. $\mathcal{N}(\text{mean, std})$ is the normal distribution with mean and standard deviation of mean and std, respectively.}
\end{center}
\end{table*}

\subsection{Event selection}
\label{sec:selection}

We follow similar criteria as \citet{Kaczmarek2026} for selecting detectable astrometric events. We apply an $F146$ $<$ 23 cut for the magnitude at baseline (in the case of luminous lenses, we consider joint source and lens flux) to exclude sources impacted by background noise. We select only events where the closest approach occurs within the bounds of GBTDS: $\min(\mathcal{T}_{GBTDS}) < t_0 < \max(\mathcal{T}_{GBTDS})$.
We also require that there is a detectable change in astrometric signal within the GBTDS time baseline,
$\Delta_{\rm ast} > \sigma_{\rm ast}/\sqrt{N}$,
where
\begin{equation}
    \Delta_{\rm ast} = \max |\vec{\delta}(t_i) - \vec{\delta}(t_j)|: t_i, t_j \in \mathcal{T}_{GBTDS}
    \label{eq:dect_change}
\end{equation}

and $N = 119$ for stacking observations taken within 24 hours under a 12.1-minute cadence. We use Eqs. (\ref{eq:full_shift}), (\ref{eq:lens_source_track}) to calculate $\vec{\delta}(t)$; we omit the parallax motion component in Eq. (\ref{eq:lens_source_track}), assuming its negligible effect on astrometric signal detectability. In the case of luminous lenses, $\vec{\delta}(t)$ is calculated including corrections for the blending effect, using relations from \citet{Dominik2000}.

We differ from \citet{Kaczmarek2026} in the definition of $\sigma_{\rm ast}$, as a new detailed study of \romanst 's position uncertainties has recently been published by \citet{McKinnon2026}, hereafter \hyperlink{cite.McKinnon2026}{McK26}. We use the data and codes published in the repository\footnote{\texttt{\url{https://github.com/KevinMcK95/gaia\_roman\_astrometry}}} accompanying \hyperlink{cite.McKinnon2026}{McK26} to define the single-exposure position uncertainty $\sigma_{\rm ast}$. We follow \hyperlink{cite.McKinnon2026}{McK26} in adding in quadrature a noise floor of 1\% of the pixel width; we use simulated data for 66-second exposures in the middle of the 5-field strip in the GBTDS footprint (see Fig.~\ref{fig:footprint}).

Finally, we require little to no blending -- i.e., at least $0.8$ of the total (source, lens, and neighbours) blended flux coming from the source, according to the \texttt{PopSyCLE}-reported $H$-band blending parameters. In addition to excluding events with a diluted astrometric signal due to blending, this cut ensures that the event is simple to model astrometrically \citep[see e.g.,][]{Bramich2018}. We also add a photometric detectability flag to our dataset, but do not use it as a selection criterion. This flag is defined as in \citet{Kaczmarek2026} and analogously to Eq. (\ref{eq:dect_change}): we require $|u_0|<2$ (corresponding to a photometric signal of $\Delta m = 0.06$ at peak) and $\Delta_{phot} > \sigma_{phot}/\sqrt{N}$, i.e., a change in magnitude exceeding the magnitude-dependent precision limit determined using Eqs. (6--7) from \citet{Kaczmarek2026}.

\subsection{Constraining event parameters}

We create simulated observations over $\mathcal{T}_{\rm GBTDS}$ for all astrometrically detectable events that have BH lenses. We simulate the ground truth positions and fluxes from Eqs.~\ref{eq:full_shift}--\ref{eq:ampl} and add white Gaussian noise with standard deviations $\sigma_{ast,phot}/\sqrt{119}$ (1-day stacking), consistent with the photometric and astrometric selection criteria described in Section \ref{sec:selection}. In addition to the nominal $\mathcal{T}_{\rm GBTDS}$, following \citet{RDox}\footnote{\url{https://roman-docs.stsci.edu/roman-community-defined-surveys/galactic-bulge-time-domain-survey}}, we add gap-filling observations with a cadence of 5 days and added noise of standard deviation $\sigma_{ast,phot}$ (single exposure). The error bars match the standard deviations of the added noise.

The model parameter priors are detailed in Table \ref{tab:constraints_priors}. Parallax factors and standard five-parameter source tracks were calculated using \texttt{astromet}\footnote{\url{https://github.com/zpenoyre/astromet.py}}. We fit the astrometric microlensing signal jointly with the standard five-parameter source astrometric solution for the unlensed source path. The fit is performed by assuming a white Gaussian noise likelihood. We use the nested sampling algorithm \citep{Higson2019} implemented by \texttt{DYNESTY} \citep{dynesty} because of its ability to find multiple modes of the posterior that are common in microlensing event modeling \citep[e.g.,][]{Kaczmarek2022}. We use random-walk sampling with multiple bounding ellipsoids and $1000$ initial live points. We use a stopping criterion based on the remaining fractional evidence of 0.01 and allocate all weight to posterior computation. For each event, we model two scenarios: with and without photometry.

\section{Results}
\label{sec:results}

\subsection{Event yields}

\begin{figure*}[t]
    \centering
    \includegraphics[width=\linewidth]{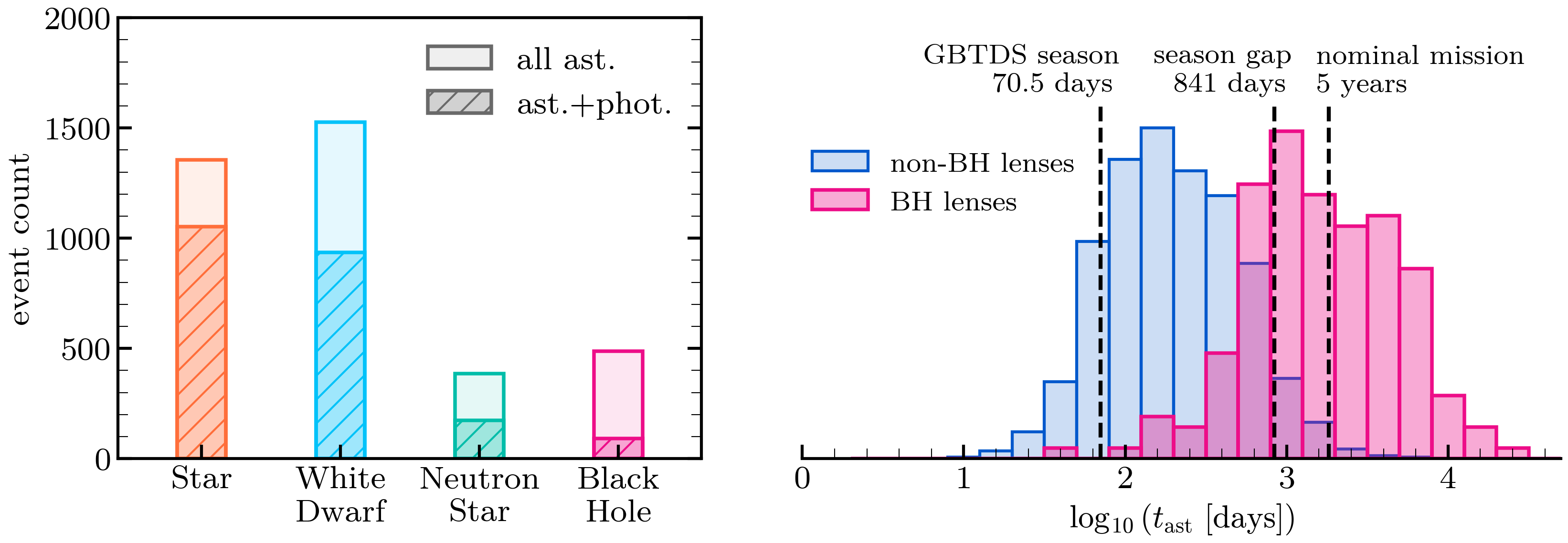}
    \caption{Left: Bar chart showing the predicted number of each lens type that will produce detectable astrometric signals during the nominal 5-year GBTDS. Hatched bars represent subsets of astrometric events that are also photometrically detectable. Right: Distributions of astrometric event timescales ($t_{\rm ast}$; Eq. \ref{eq:tast}) for the sample of detected events with non-BH (blue) and BH (magenta) lenses. Each distribution is normalized and therefore contains no information on the relative number of BH and non-BH events. Marked as vertical dashed lines are 70.5 days, 841 days, and 5 years, which correspond to the duration of a single season, large season gap (between seasons 3 and 4), and total survey duration for the nominal GBTDS \citep{Penny2019}. See Section \ref{sec:characteristics} for discussion.}
    \label{fig:yield_timescale}
\end{figure*}

The total yields of selected events within the GBTDS footprint are shown in Fig.~\ref{fig:footprint}. We predict a total of 3755 $\pm$ 103 astrometric events over the full duration and area of GBTDS, 487 $\pm$ 37 of these have BH lenses.  Of these events, 1540 $\pm$ 66 do not pass photometric detectability ($|u_0|<2$ and $\max (\Delta_{phot}) > \sigma_{phot} / \sqrt{N}$; Section \ref{sec:selection}) cuts; 398  $\pm$ 33 of which have BH lenses (see Fig.~\ref{fig:yield_timescale}, left panel).
Therefore, we find that including astrometry-only events increases the yield of detectable BH lenses with astrometric signals by a factor of 5.4. This increased BH yield does not necessarily imply a proportional increase in overall population constraints. This is because mass measurements are more challenging and are likely to have higher uncertainties in the absence of a photometric signal (see Section \ref{sec:constraint_results}). 

We also find that astrometry-only events are an efficient channel for BH detection. $25.8\%$ of events with only astrometric signals detectable are caused by BH lenses. This is large compared with the corresponding BH lens proportions of $4.0\%$ and $1.1\%$ of events detectable in photometry and astrometry, and photometry only, respectively. This is due to a combination of the astrometric signals' dependence on the lens mass, which is typically higher for BHs compared with other lenses, and the astrometric selection criteria that were applied. $97\%$ of the photometrically detectable BH lenses are also astrometrically detectable.

\subsection{Characteristics of the astrometric signals}\label{sec:characteristics}

\begin{figure*}
    \centering
    \includegraphics[width=\textwidth]{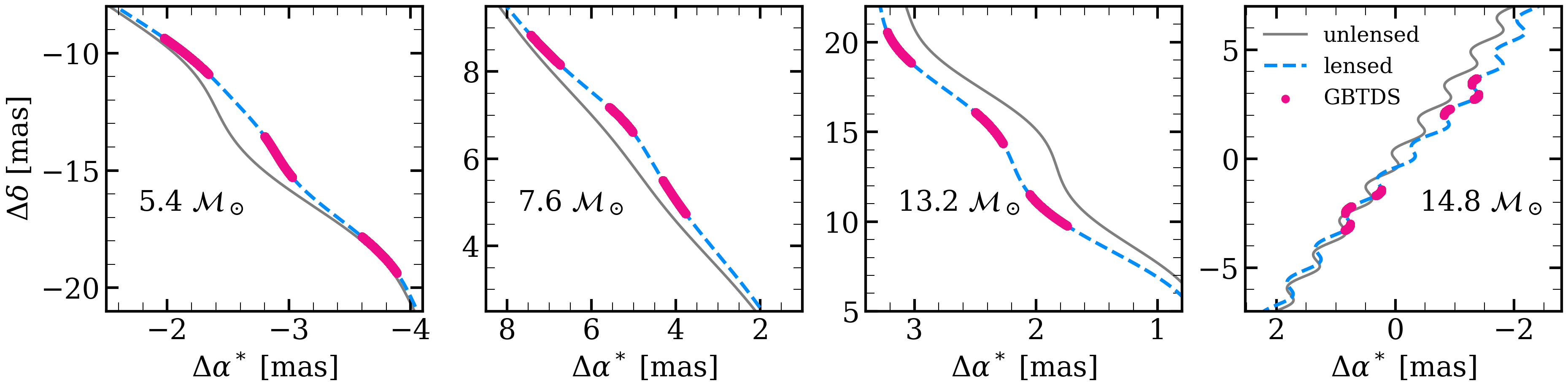}
    \caption{Example black hole astrometric microlensing event trajectories that will be detectable during the GBTDS (with respect to a reference position at the midpoint of the survey). Grey solid lines show the unlensed background source trajectory. Blue dashed lines show the lensed observed background source trajectory.  Magenta circles represent GBTDS observations. The clusters of measurements are the assumed nominal GBTDS 70.5-day observing seasons. The low-cadence gap-filling observations will have significantly higher uncertainties than stacked observations in high-cadence seasons, and therefore are not shown.}
    \label{fig:trajectories}
\end{figure*}

The right panel of Fig. \ref{fig:yield_timescale} shows the distribution of astrometric event timescales for BH and non-BH lenses that are detectable by the GBTDS. At this point, it is useful to reiterate our astrometric detection criteria. We require a detectable \textit{change} in astrometric deflection within the GBTDS survey (Eq. \ref{eq:dect_change}). This criterion excludes events that could have a detectable absolute astrometric signal but do not vary significantly enough within the GBTDS time baseline to be detectable \citep[e.g.,][]{Dominik2000}.

For events that pass the astrometric detection criteria, Fig. \ref{fig:yield_timescale} shows that the timescale distribution peaks around 1000 days, with many BH events having astrometric timescales comparable to the total primary mission time. For these events, the GBTDS will only detect part of the full astrometric microlensing event, and the source star will be displaced astrometrically for the full GBTDS. This means that for the majority of BH events, multiple seasons, if not all seasons, of the GBTDS will be required to characterize the full astrometric signal. Fig.~\ref{fig:trajectories} shows example trajectories of different events with BH lenses whose astrometric signals will be detectable during the GBTDS. Most events span multiple GBTDS seasons of data. This also suggests that any gap-filling observations will increase the number of data points on the detectable signal for the majority of BH events.

Fig.~\ref{fig:yield_timescale} also shows that there is a minority of short-timescale events with $t_{\rm ast}< 841$ days, suggesting that gap-filling observations between seasons 3 and 4 are likely to boost yields of events that would otherwise fall completely within this gap. Finally, Fig.~\ref{fig:yield_timescale} shows that detectable events with BH lenses that have an astrometric timescale less than the duration of a single high-cadence season, 70.5 days, are extremely rare. This suggests that if an astrometric microlensing signal is completely contained within one GBTDS season, we find that it is unlikely that it is caused by a BH.

\subsection{Lens constraints}\label{sec:constraint_results}

\begin{figure*}[t]
    \centering
    \includegraphics[width=1.0\linewidth]{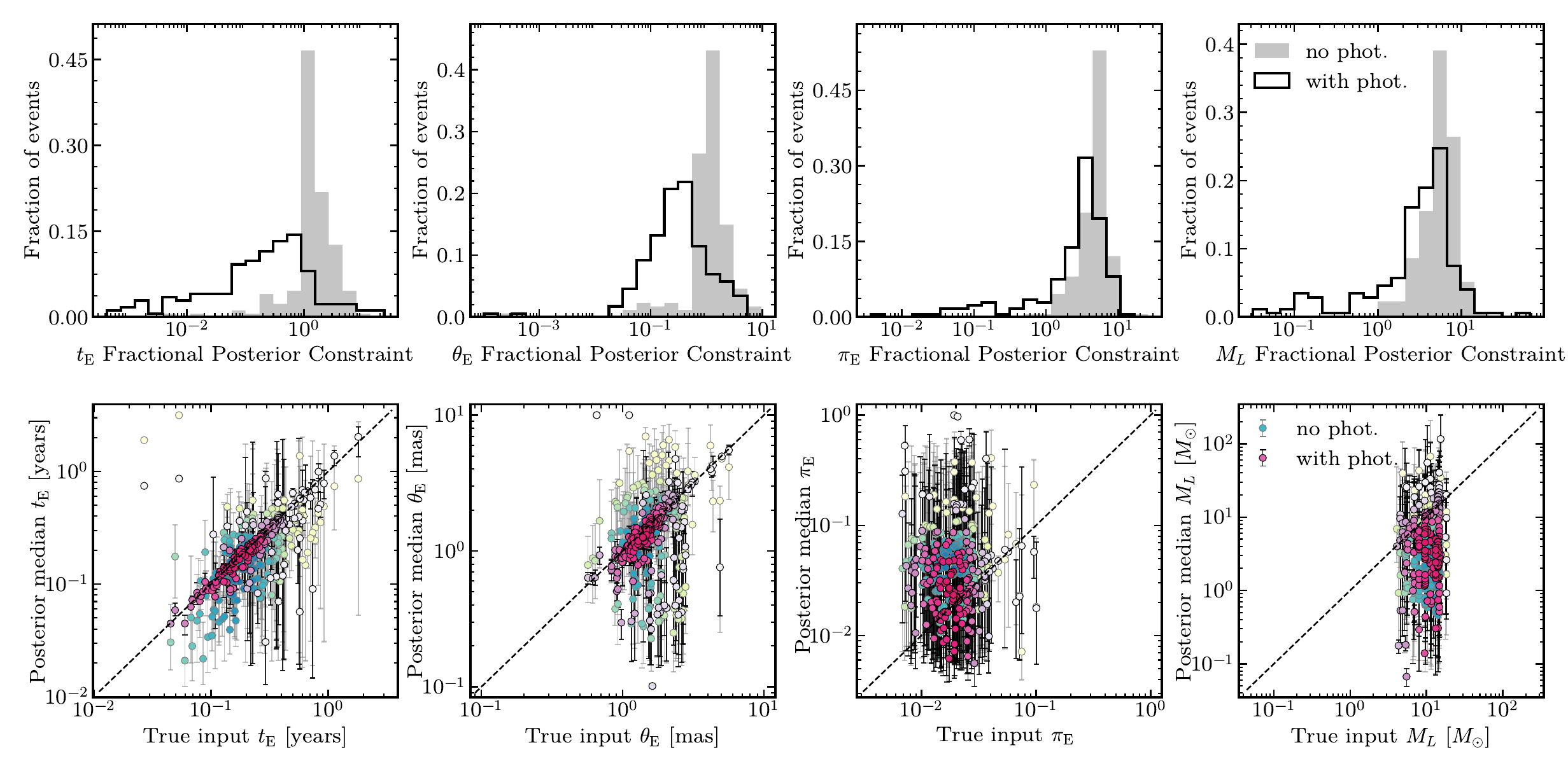}
    \caption{Posterior precision and accuracy of the microlensing event parameters and the corresponding lens mass for the detected sample of events. Top panels: Fractional posterior constraint is defined as the (84$^\text{th}$--16$^\text{th}$)/(50$^\text{th}$) percentile of the posterior distribution. Including the photometry in the model fitting improves constraints for all parameters. $\theta_{\rm E}$ and $t_{\rm E}$ are typically better constrained than $\pi_{\rm E}$ and $M_{L}$. Bottom panels: Fitted versus true input values for event parameters. For clarity, data points are coloured by density estimated with a 2D Gaussian KDE, with darker shades (of magenta and green for fits with and without photometry,  respectively) signifying higher point density. Similar to the posterior precision in the top panels, $\theta_{\rm E}$ and $t_{\rm E}$ are typically more accurately recovered than $\pi_{\rm E}$ and $M_{L}$. Including the photometry in the model also improves posterior accuracy.}
    \label{fig:constraints}
\end{figure*}

For the set of detectable events we find that including (and excluding) the photometry, $\theta_{\rm E}$ and $t_{\rm E}$ can be constrained with a typical (median) precision of $\approx30\%$ ($110$\%) and $\approx25\%$ ($135\%$). $\pi_{\rm E}$ and consequently $M_{L}$ are less well constrained with typical precisions of $\approx 340\%$ $(520\%)$ and $\approx 350\%$ $(580\%)$, respectively. In particular, high $\pi_{\rm E}$ introduces greater freedom in the modelled trajectory, and the fit may absorb random noise or shifts from linear motion into an artificially inflated parallax signal. Therefore, $\pi_{\rm E}$ is often overestimated, resulting in underestimated masses, especially at low signal-to-noise ratios. Fig. \ref{fig:constraints} shows the distribution of these constraints across the detected sample and their accuracy. Similarly to the posterior precision, the posterior accuracy is better for $\theta_{\rm E}$ and $t_{\rm E}$, compared with $\pi_{\rm E}$ and $M_{L}$. We note that, for a subset of low signal-to-noise events, multiple or extended posterior modes were found. We defer analysis of these degeneracies to future work. All posterior constraints are reported including all modes.

Including the photometry, even when the photometric microlensing signal amplitude is low or undetectable, improves the constraints on all parameters. This is even the case for $\theta_{\rm E}$ -- a parameter for which only the astrometric signal contains direct information. This is because the amplitude, or lack thereof, of the photometric signal places a lower bound on $u_{0}$ and/or $|\boldsymbol u(t)|$. This bounding of the normalized lens--source separation allows a tighter constraint on $\theta_{\rm E}$ because the astrometric signal is a function of both $|\boldsymbol u(t)|$ and $\theta_{\rm E}$ (see Eq. \ref{eq:full_shift}). Similar reasoning on bounding $|\boldsymbol{u}(t)|$ can explain the increase in precision for $t_{\rm E}$ and $\pi_{\rm E}$ when including photometry. Additionally, even at low signal to noise, the photometry can contain direct information on $t_{\rm E}$ and $\pi_{\rm E}$ which will also act to increase posterior precision when photometry is included.

Better constraints for $\theta_{\rm E}$ and $t_{\rm E}$, compared with $\pi_{\rm E}$ and consequently $M_{L}$, are expected \citep[e.g.,][]{Fardeen2023, Kaczmarek2026b}. This is because, to first order, $\theta_{\rm E}$ and $t_{\rm E}$ are correlated with the amplitude and duration of the astrometric signal, respectively. These quantities are usually straightforward to constrain. In contrast, $\pi_{\rm E}$ typically produces smaller deviations in the astrometric microlensing signal. This is because the astrometric signal is not sensitive to small lens--source parallax-induced |$\boldsymbol{u}$| deviations at large (|$\boldsymbol{u}$|>2) separations \citep{Dominik2000}.

Fig. \ref{fig:constraints} shows that while there is a heavy left tail of $\approx 20\%$ of events where we can obtain $M_{L}$ posterior constraints between $1\%$ and $100\%$, the majority of these events will not yield a tight constraint on $M_{L}$. This unfortunate situation suggests that these events will not lead to impactful constraints on the shape of the BH mass function. However, in the context of a Galactic model, these events could still provide constraining information on the number of isolated BHs. In the following Section we turn our attention to exploring this possibility. That is, we ask the question -- if we cannot constrain $M_{L}$ from the event, can we still identify the lens as a BH?

\subsection{Lens classification}
\label{sec:lens_class}

\begin{figure*}[t]
    \centering
    \includegraphics[width=\linewidth]{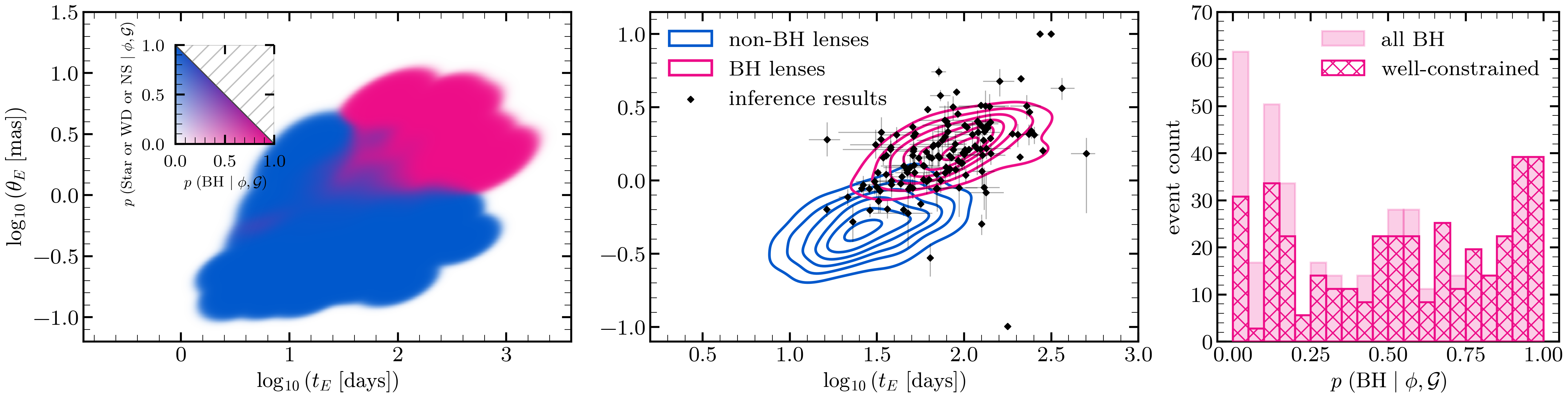}
    \caption{Left: The space of high-signal parameters, $\log_{10} t_{\rm E}$--$\log_{10} \theta_{\rm E}$, coloured by probabilities of belonging to a given lens class $p (\text{class}_L | \phi, \mathcal{G}) $. Non-BH astrophysical lens classes are grouped together, effectively: {BH, Star or WD or NS, None}. Each point $\phi$ on a $1000\times1000$ grid is coloured by the probability of belonging to either of the astrophysical classes (magenta -- BH, blue -- non-BH). The inset represents a 2D colorbar; probabilities are normalized to 1, making the space above the diagonal where $\sum_{i} p(\text{class}_{L,i}) > 1$ forbidden. There is a distinct region (which can be conservatively approximated by cuts of $t_{\rm E}>100$ days and $\theta_{\rm E}>1.2$ mas) where the probability of the lens being a BH is $\approx$1. The BH and non-BH regions are well-separated, with a relatively small region belonging to the overlap (purple) where confusion between BH and non-BH classes is high. Center: Distributions of BH and non-BH lenses in the $\log_{10} t_{\rm E}$--$\log_{10} \theta_{\rm E}$ space, along with inference results for the mock dataset of BH-lens astrometric events (Section~\ref{sec:constraint_results}). Contours represent the simulated astrometric events caused by black holes (magenta) vs. all other types of lenses: stars, WDs, and NSs (blue); they were built using a Gaussian kernel density estimate and contain (80, 65, 50, 35, 20, 5)\% of the simulation samples. Black data points represent astrometric BH events that are well-constrained, i.e., that have both $t_{\rm E}$ and $\theta_{\rm E}$ fractional posterior constraints $<1$; this subset contains 79\% of our simulated astrometric BH events. The position in the plot represents median $t_{\rm E}$ and $\theta_{\rm E}$ values from samples of the nested sampling fit to the mock GBTDS observations; error bars represent 16$^{\text{th}}$--84$^{\text{th}}$ percentiles from samples. Right: histogram of $p (\text{BH} | \phi, \mathcal{G}) $ -- the probability of the lens being a black hole based on its posterior $\phi$ and the Galactic model $\mathcal{G}$ -- as returned by \texttt{popclass} for all simulated astrometric BH events (solid filled) and the well-constrained subset (cross-hatched). See Section~\ref{sec:lens_class} for detailed discussion of the \texttt{popclass} classification results.}
    \label{fig:bh_tE_thetaE}
\end{figure*}

Fig.~\ref{fig:bh_tE_thetaE} shows where the population of BH lenses is located in the $t_{\rm E}$--$\theta_{\rm E}$ space relative to the population of non-BH lenses. Analyzing the separation of lens types in microlensing observables is important because it could be used to identify high-confidence BH candidates or, at least, provide a lens class probability that can be marginalized over in population model inference \citep{Perkins2023,Perkins2026,Sallaberry2025}.

BH lenses typically have larger $t_{\rm E}$ and $\theta_{\rm E}$ than the other lens types. Although the BH and non-BH lens populations are mostly separated in the $t_{\rm E}$--$\theta_{\rm E}$ space, there is overlap. Moreover, the BH lens population is situated in the tails of the much more abundant populations of stellar and WD lenses. One notable feature of the $t_{\rm E}$--$\theta_{\rm E}$ space shown in Fig.~\ref{fig:bh_tE_thetaE} is that the BH lenses are located in a region of high signal compared with the non-BH population. That is, BH lens events are more likely to have a larger $\theta_{\rm E}$ and, therefore, a larger astrometric signal. This suggests that $\theta_{\rm E}$, and therefore their location and lens class, in $t_{\rm E}$--$\theta_{\rm E}$ space can be measured more precisely. This is in contrast to photometric microlensing events monitored from the ground, where $\pi_{\rm E}$ and $t_{\rm E}$ can be measured, but BH lenses are located in a low-signal region of large $t_{\rm E}$ but small $\pi_{\rm E}$, resulting in uncertain lens classifications \citep[e.g.,][]{Lam2020,Kaczmarek2025}. The left panel of Fig.\ref{fig:bh_tE_thetaE} suggests that there is a region of $t_{\rm E}$--$\theta_{\rm E}$ space ($t_{\rm E}>100$ days and $\theta_{\rm E}>1.2$ mas), populated with detectable astrometric events, where the probability of a BH lens is close to $100\%$.

To predict how well BH lenses can be identified with GBTDS data, we classify our simulated BH events using \texttt{popclass}\footnote{\url{https://github.com/llnl/popclass}} \citep{Sallaberry2025}. The classification framework calculates $p (\text{class}_L | \phi, \mathcal{G}) $ -- the probability of the lens belonging to a given lens class based on its posterior $\phi$ and the Galactic model $\mathcal{G}$. We use posteriors from joint modelling of photometric and astrometric time series for classification, as they provide more precise parameter constraints (Fig.~\ref{fig:constraints}). We construct the astrophysical lens classes (Star, White Dwarf, Neutron Star, Black Hole) on the basis of all astrometrically detectable events from our simulation; we also construct a None class that captures combinations of parameters that are inconsistent with astrophysical lenses in the Galactic model. We refer to \citet{Sallaberry2025} for a detailed discussion of the classification procedure, and to \citet{Kaczmarek2026} for the construction of the None class over the $\log_{10} t_{\rm E}$--$\log_{10} \theta_{\rm E}$ space which we reproduce here.

Fig.~\ref{fig:bh_tE_thetaE} shows the classification results. In the center panel, we show all BH events that are well-constrained with the nested sampling fits to mock GBTDS data. We consider an event well-constrained if it has fractional posterior constraints on both $t_{\rm E}$ and $\theta_{\rm E}$ lower than $1$. We find that most of the well-constrained fits are located in the parameter space region occupied by BH lenses. 56\% of all events (and 64\% of the well-constrained sample) are classified as Black Hole (i.e., $p(\text{class}_L | \phi, \mathcal{G})$ is the highest for $\text{class}_L = \text{BH}$). 3.4\% (0.7\%) are assigned the Star class, 1.1\% (1.4\%) -- the Neutron Star class, and 1.7\% (1.2\%) -- the None class. The most significant source of confusion by far, at 37\% (31\%), is the White Dwarf class. \citet{Kaczmarek2025} have noted that high-probability regions for WD and BH lenses in the observable parameter space are close, giving rise to sharp $p(\text{class}_L | \phi, \mathcal{G})$ gradients where ambiguous classifications are likely to occur (including the only known isolated BH, OB110462, yielding a high WD probability). To improve ambiguous WD/BH classifications, parameter constraints must be tightened (e.g., by extending \textit{Roman}'s observing baseline beyond the nominal mission or incorporating external data).

We also report Spearman's correlation \citep{Spearman1904} between event parameters and $p(\mathrm{BH})$ as a simple gauge of which events are most likely to be recovered. $p(\mathrm{BH})$ is most correlated with astrometric signal-to-noise ratio SNR$_{ast}$ ($\rho = 0.66$) and $\theta_E$ ($\rho = 0.54$); it is moderately ($\rho \approx 0.4$) correlated with $\pi_{\rm E}$ (which does not meaningfully contribute to astrometric signal and classification, but is a function of $\theta_E$) and UWE; correlation with other parameters ($t_{\rm E}, \ \text{SNR}_{phot}, \  |u_0|$) is weak ($\rho < 0.35$). Interestingly, though there are clear linear trends between $\log (\text{excess UWE})$ and $p(\mathrm{BH})$ or ($\theta_{\rm E}, \ t_{\rm E}$) fractional constraints, events with $ \text{UWE}\lesssim 1$ also occur at all $p(\mathrm{BH})$ ranges and can have well-constrained posteriors. Those events should be identified with other metrics, e.g., taking into account the variability of excess signal in time.

\subsection{Extracting events from the GBTDS}

So far, we have investigated the sample of astrometric microlensing events that can be detected above some detection threshold. This represents an idealized scenario in which we are sure the source in question is undergoing an astrometric microlensing event. In practice, we may not know \textit{a priori} if this is the case. We must separate astrometric microlensing events from sources with no motion residual relative to the standard proper motion and parallax solution and from sources undergoing accelerations due to different effects \citep[e.g., astrometric binaries;][]{Lam2026}. In this section, we deal with the former and defer the latter issue to future work.

We have performed the 5-parameter ($\alpha, \delta, \mu_{\alpha*}, \mu_\delta, \pi$) astrometric fits for all our simulated astrometric events and calculated unit weight error (UWE) values of the fits using the \texttt{astromet}\footnote{\url{https://github.com/zpenoyre/astromet.py}} package. The expected value of UWE for deviations from a 5-parameter track fully explainable by random noise is 1; high excess UWE values point towards extra astrometric signals, such as astrometric microlensing shifts. A similar treatment, using renormalised UWE (RUWE), is commonly applied to extract anomalous astrometric fits in the \textit{Gaia} mission data, as it provides a single, easily interpretable value that has been extensively calibrated with real data and simulations \citep[e.g.,][]{Lindegren2018, Lindegren2021, Penoyre2022, CastroGinard2024} and has been proposed for finding astrometric signals of \textit{Gaia}'s microlensing events \citep{Jablonska2022, Kaczmarek2026b}. We use UWE as a proxy for RUWE, expecting its renormalisation as a function of magnitude and colour to be performed over the course of \textit{Roman}'s operations.

\begin{figure}
    \centering
    \includegraphics[width=\linewidth]{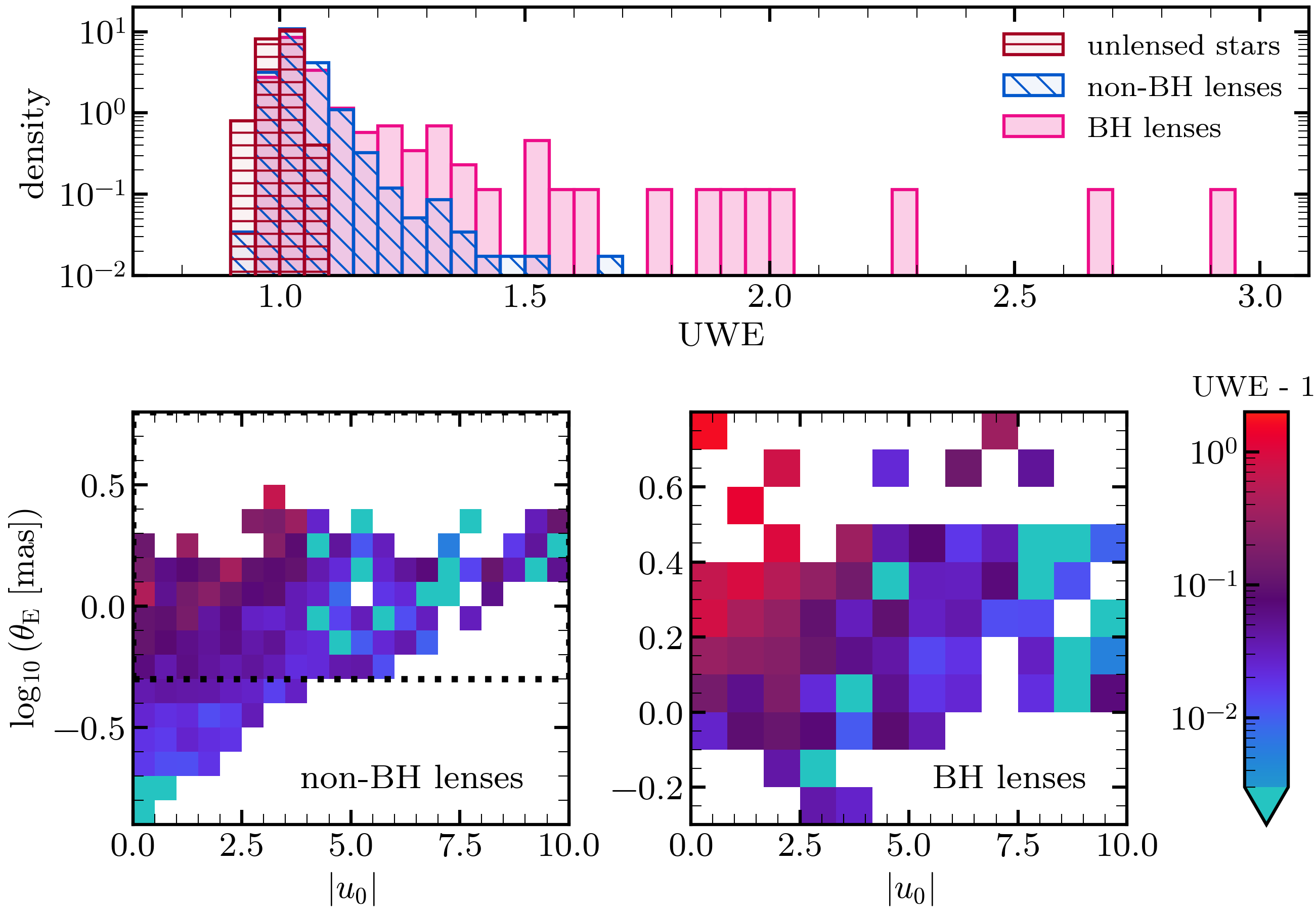}
    \caption{Top: Unit weight error (UWE) distributions for a sample of unlensed stars (red, horizontal hatch) and lensing events with non-BH lenses (blue, left-leaning hatch) and BH lenses (magenta, solid filled). Astrometric lensing introduces a tail of high-UWE tracks, which is especially prominent for BH lenses. Bottom: Excess UWE as a function of impact parameter (absolute value, $|u_0|$) and $\theta_{\rm E}$ for events with non-BH (left) and BH (right) lenses. As BHs occupy a distinct region of parameter space (see Fig.~\ref{fig:bh_tE_thetaE}), the right plot has smaller $\theta_{\rm E}$ extents, marked by the dotted contour within the left plot. Mean values of UWE - 1 are plotted within 2D bins; redder colours signify higher UWE excess, while negligible or negative values are plotted with turquoise. BH lenses have a significantly higher distribution of excess UWE and are responsible for all of the events with extreme ($>$ 0.8) values. While the highest excess (beyond 5-parameter tracks) astrometric signal is observed for events that also have high photometric signal (low $|u_0|$), there is a notable population of events that have a significant excess UWE at high ($|u_0|>$ 2) impact parameters; these events do not have typical microlensing light curves with prominent peaks and are unlikely to be picked by microlensing detection pipelines.}
    \label{fig:uwe_map}
\end{figure}

Fig.~\ref{fig:uwe_map} (top) shows the UWE distribution with respect to the standard 5-parameter astrometric solution, positions, proper motion, and parallax, for unlensed sources, sources lensed by non-BHs, and sources lensed by BHs. As expected, the population of lensed sources causes a heavy right tail in the UWE distribution that separates from the unlensed source population at UWE $\approx1.1$. The population of sources lensed by BHs shows an even heavier and flatter tail, with extreme values of UWE $>2$. We find no non-BH lenses with UWE $>2$. Assuming other sources of astrometric accelerations, e.g., binaries or systematics, are removed, extreme values of UWE point to a likely BH lens.

Fig.~\ref{fig:uwe_map} (bottom) shows how UWE varies for different values of $|u_{0}|$ and $\theta_{\rm E}$. These event parameters control the amplitude and shape of the astrometric signals. A larger $\theta_{\rm E}$ corresponds to a larger astrometric amplitude for an event with a given $|u_{0}|>\sqrt{2}$. Smaller $|u_{0}|$ values correspond to more peaked astrometric signals. Fig.~\ref{fig:uwe_map} (bottom, right panel) shows that BH lenses with the smallest $|u_{0}|$ values (most peaked signals) and larger $\theta_{\rm E}$ values (largest amplitudes) have the highest excess UWE values. These events have the most temporally localized and largest deviations from the standard astrometric solutions.

With $\theta_{\rm E}$ held constant, as $|u_{0}|$ increases, UWE tends to decrease for both BH and non-BH lenses. This is because larger $|u_{0}|$ values lead to less peaked deviations, spreading the signal over a greater timescale. These events are more likely to have a lower UWE because the astrometric signal is more likely to be mistaken for a different set of standard astrometric solution parameters. Notwithstanding the lower UWEs for larger $|u_{0}|$, there is a notable population of events that have a significant excess UWE at high ($|u_0|>2$) impact parameters. These events will not have typical microlensing light curves with prominent peaks and therefore are less likely to be identified by the MSOS detection pipeline.

\subsection{The astrometric sensitivity function}

We define the sensitivity to astrometric signal as the fraction $N_{\rm BH, ast}/N_{\rm BH, tot}$; $N_{\rm BH, ast}$ denotes the number of events with BH lenses that have astrometric signals detectable over the nominal GBTDS in a given sample of simulation results, and $N_{\rm BH, tot}$ -- the total number of close ($|u_0| <10$) BH lens -- background source approaches in the sample. To investigate how the sensitivity varies across event parameter space, we calculate this fraction separately in bins of event parameters (lens mass, event timescale, etc.).

The sensitivity estimates were impacted by significant noise due to the relatively low numbers of detectable events per bin. Astrometric detectability is dependent on parameters that are not physically meaningful, $u_0$ and $t_0$. Namely, events with lower $|u_0|$ cause higher astrometric signals (see e.g. Fig.~\ref{fig:uwe_map}, Eq.~\ref{eq:full_shift}); events with $t_0$ during the large season gap (see e.g. Fig.~\ref{fig:trajectories}) may have no observation epochs covering the high-signal part of the astrometric shift, and are missed more often. To alleviate this effect, we created 50 copies of each simulated BH lens event, for which all parameters except $u_0$ and $t_0$ are unchanged; for the 49 new instances, $u_0$ and $t_0$ are replaced with random values allowed by the simulation bounds: $u_0 \in \mathcal{U}(-10, 10); t_0 \in \mathcal{U}(\min{\tau_{GBTDS}}, \max{\tau_{GBTDS}})$.

We estimate uncertainties on the detection sensitivity using binomial proportion confidence intervals for selecting $K$ detectable events out of $N$ total simulated lens--source approaches. To account for the uncertainties decreasing when the impact of nuisance parameters ($u_0$, $t_0$) is minimized, we calculate the effective sample size $n_{\rm eff}$ for each 50-event ensemble using intra-class correlation \citep[ICC;][]{Kish1965} for binary detection results ($d=0$ -- not detectable or $d=1$ -- detectable). The effective sample size of a bin is then $N_{ \rm eff} = \sum_{i=1}^{N_{\rm BH,tot}}{n_{\rm eff, i}}$ (summed over all unique BH lenses within each bin); sensitivity in the bin is $S = \frac{1}{N_{\rm BH,tot}}\sum_{i=1}^{N_{\rm BH,tot}}\bar{d_i}$, where $\bar{d_i}$ is the mean $d$ for the $i$-th set of 50 `copied' events with the same BH lens. We report Wilson intervals \citep{Wilson1927} using sensitivity of $S$ within a sample size of $N_{\rm eff}$ for each bin. The effective sample sizes $N_{\rm eff} \ll 50N$ and $N_{\rm eff} \approx N$, primarily because a high fraction of events get rejected at the early stages due to magnitude and blending cuts (Section \ref{sec:selection}) which are independent of ($u_0$, $t_0$). Additionally, we find that events with high BH lens masses have low sensitivity to ($u_0$, $t_0$) changes and return particularly high ICC values. However, copying events with ($u_0$, $t_0$) replacement brings a notable decrease of noise on the low-$M_L$ edge, where astrometric signals have shorter timescales and detectability is strongly dependent on $u_0$. 

\begin{figure*}[t]
    \centering
    \includegraphics[width=\linewidth]{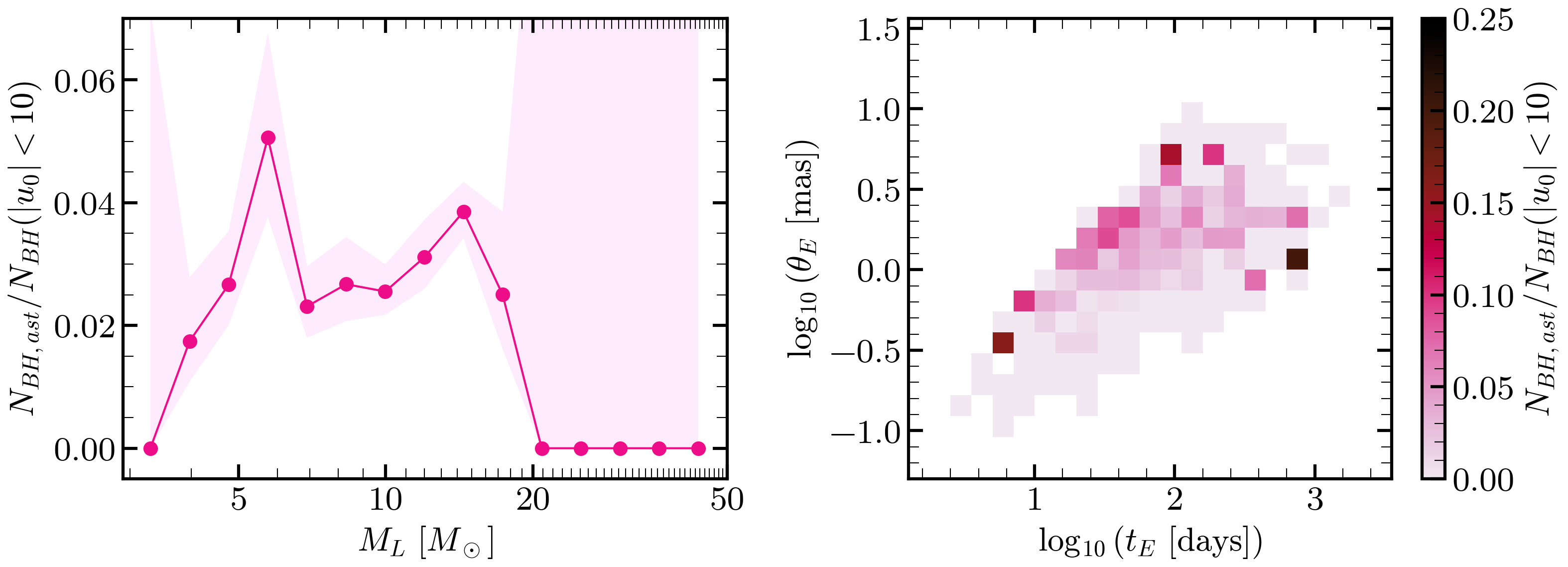}
    \caption{BH lens detection sensitivity as a function of mass (left) and over the joint $t_{\rm E}-\theta_{\rm E}$ observable space (right). The left panel shows that the GBTDS is sensitive through astrometric microlensing to BHs with masses between $\approx3-19 M_{\odot}$. BHs on the low-mass end are less likely to be detected due to their typically smaller $\theta_{\rm E}$ and therefore smaller astrometric signals. BHs with masses of $\approx5-6 M_{\odot}$ are most likely to be detected. The sensitivity function increases again for BH masses of $\gtrsim10 M_{\odot}$, reaching a secondary peak at $\approx14-15 M_{\odot}$. This peak marks the transition of median $t_{\rm ast}$ per bin becoming longer than the nominal GBTDS time. Events caused by more massive BHs have high $\theta_{\rm E}$, but a low rate of change of the astrometric shift; to detect those events, additional observations long before or after the nominal surveys would be needed to establish the astrometric baseline for the source. For all bins with $M_{L} > 19 M_{\odot}$, the estimated uncertainty intervals (Wilson intervals for $k=0$ successes) per bin are very large. However, we note that we consistently register no detections of any BH lens with $M_{L} > 18.3 M_{\odot}$ within our simulated dataset. The right panel shows that events with large $\theta_{\rm E}$ and intermediate $t_{\rm E}$ have the highest detection efficiency. These events typically peak and return to baseline within the GBTDS. Detectable events span a wide range of $t_{\rm E}$; within each $t_{\rm E}$ bin, the highest possible $\theta_{\rm E}$ is strongly preferred for detectability. }
    \label{fig:selection}
\end{figure*}

In Fig.~\ref{fig:selection}, we present detection sensitivity as a function of mass and location in the $t_{\rm E} - \theta_{\rm E}$ space. We find that the GBTDS is sensitive to BH masses of $\approx 3-19M_\odot$. Sensitivity is highest for $5-6M_\odot$ BHs, whose astrometric signals are rapidly changing and largely contained within the GBTDS timespan. In the $t_{\rm E} - \theta_{\rm E}$ space, astrometric signals can be detectable across a wide range of timescales, with intermediate timescales being most preferred; for each $t_{\rm E}$ range, the highest astrophysical $\theta_{\rm E}$ values are strongly preferred. The high-mass end of the sensitivity function is burdened by high statistical uncertainties, as BHs with masses $>20 M_{\odot}$ are rare in the adopted Galactic model \citep[e.g.,][]{Rose2022}.

\section{Discussion and conclusions}
\label{sec:discussion}

We simulated astrometric microlensing in the five-year Roman GBTDS, extending the search to events with impact parameters $|u_0|<10$. Under our adopted Galactic model and selection criteria, we predict $3755\pm103$ events with detectable astrometric signals, $487\pm37$ of which are events caused by isolated BHs. Of these, $1540\pm66$ events, including $398\pm33$ BH events, do not pass our photometric detectability cuts. Therefore, including these astrometry-only events increases the predicted sample of BH lenses with detectable astrometric signals by
a factor of $5.5$ relative to the photometrically detectable BH sample. This demonstrates the importance of extending GBTDS microlensing searches beyond events identified through their light curves. We find that the GBTDS survey will be most sensitive to BHs of $5-6 M_\odot$, and not sensitive to BHs with high ($>19M_\odot$) masses. Astrometric detection fraction is highest for intermediate timescales and high angular ($\theta_{\rm E}$) scales.

\subsection{Astrometric detection and population constraints}

Astrometry-only events are promising for finding isolated BHs. BH lenses account for $25.8$\% of this sample, compared with 4.0\% of events detectable in both photometry and astrometry. This difference reflects both the larger $\theta_{\rm E}$ of BH lenses and our selection criteria. Astrometric searches therefore offer not only a larger potential BH sample, but also a sample less dominated by stellar lenses. Our results complement forecasts focused on events with large photometric signals \citep{Pruett2022,Sajadian2023} and support the importance of the large impact parameter regime originally explored for the GBTDS in \citet{Fardeen2024}.

However, detecting an astrometric signal, identifying the lens as a BH, and measuring the BH mass are distinct and potentially non-trivial tasks for a given microlensing event. Our simulations show that $\theta_{\rm E}$ and $t_{\rm E}$ are generally better constrained than $\pi_{\rm E}$ for these events. This comparatively weak constraint on $\pi_{\rm E}$ is the primary obstacle
to precise mass measurements. Consequently, this increase in detectable BH events does not translate directly into an equivalent increase in precisely measured BH masses for the GBTDS.

Photometry is valuable even for events that fail our photometric detection criteria. A low-amplitude light curve, or even the absence of measurable amplification, constrains the allowed lens--source separation and reduces degeneracies in the astrometric model. Joint modelling improves constraints even on $\theta_{\rm E}$, despite photometry containing no direct measurement of this parameter.

For this set of events, $t_{\rm E}$ and $\theta_{\rm E}$ are typically well-constrained and provide information about lens class. Within our adopted Galactic model, events with $t_{\rm E}>100$ days and $\theta_{\rm E}>1.2$ mas occupy a region with a high BH lens probability. Most events are correctly classified with the \texttt{popclass} classification software as having BH lenses, based on their mock GBTDS observations. Almost all of the misclassified events are assigned the `White Dwarf' class, owing to the proximity of WD and BH events in the observable parameter space and the high number of WD lenses.

Even events without precise individual mass measurements could constrain the abundance of isolated BHs. This analysis would need to combine event counts and parameter posteriors with a GBTDS astrometric selection function, while marginalizing uncertain classifications and Galactic model assumptions \citep[e.g.,][]{Perkins2026}. While we expect a minority of these microlensing events to yield precise BH mass measurements, the majority of this sample will not. We therefore expect this sample of microlensing events to be a promising probe of the isolated BH abundance rather than a precise probe of the shape of the isolated BH mass function.

\subsection{From detectable signals to confirmed BHs}

Our predicted yields describe an ideal sample of  detectable signals, rather than a predicted yield of recovered and confirmed BH events. Specifically, the selection criterion in Eq. \ref{eq:dect_change} requires a measurable change in the microlensing deflection. It does not require that this change remain significant after fitting the source position, proper motion, and parallax. Slowly varying microlensing signals can be somewhat absorbed into these five astrometric parameters, making them more difficult to extract in a blind search.

The UWE distributions in Fig. \ref{fig:uwe_map} show this distinction. Only $\approx18\%$ of the selected BH events have $\mathrm{UWE}>1.2$, and $\approx2\%$ have $\mathrm{UWE}>2$. Although we find no non-BH lenses above the latter threshold in our simulated sample, this does not establish a pure BH selection criterion. Astrometric binaries are the main astrophysical source of increased UWE \citep[e.g.,][]{Penoyre2022, CastroGinard2024}; unmodelled blending and instrumental systematics can also produce large residuals \cite[e.g.,][]{GaiaEDR3val, Holl2023}. On the other hand, a modest UWE does not necessarily imply that a coherent microlensing signal cannot be recovered from the
data. In particular, we find events with negligible to no excess UWE that have strong ($t_{\rm E}$, $\theta_{\rm E}$) posterior constraints and high-confidence classification as a BH.

A search strategy with actual GBTDS data could use excess astrometric residuals as an initial diagnostic, followed by joint astrometric and photometric
modeling and comparison against alternative explanations. These other explanations should include unlensed stars, astrometric binary systems, and crowded-field systematics, rather than only comparisons among different lens types.

\subsection{Limitations}

The predictions in this study are dependent on the adopted Galactic model, compact-remnant prescription, and observing noise assumptions. The initial--final mass relation determines the underlying BH mass distribution, while natal kicks affect lens kinematics and therefore event timescales. The stellar density, bulge kinematics, extinction prescription, and source luminosity distribution also affect the number of lens--source pairs \citep[e.g.,][]{Lam2020,Rose2022}. The uncertainties quoted for our event yields should not be interpreted as including these sources of model uncertainty. Exploring alternative remnant and Galactic models will be required before placing constraints on the Galactic BH population \citep[e.g.,][]{Perkins2026}.

The mass-dependent GBTDS BH sensitivity requires similar considerations. Within our simulation, sensitivity is highest for BH masses of approximately $5$--$6\,M_\odot$, and detectable BH lenses span approximately $3$--$19\,M_\odot$. We find no detections above $18.3\,M_\odot$. This absence is not a model-independent upper mass
limit for the GBTDS. Our sensitivity is defined in the context of our event detection criteria using the distances, velocities, and source properties, provided by the Galactic model. The high-mass BH bins in Fig. \ref{fig:selection} also have substantial statistical uncertainties due to the rarity of high-mass BHs under the adopted initial--final mass relation.

Our simulations and constraint results assume white Gaussian errors and a daily stacked precision obtained from 119 exposures. Realizing this gain requires sufficient independence between the astrometric errors of individual exposures. Temporally correlated errors and crowding residuals could reduce the improvement obtained by stacking, even if the single-exposure uncertainty model is accurate \citep{McKinnon2026,Lam2026}. Their impact should be assessed using the actual incoming GBTDS observations.

The detection criteria that the source star contributes at least 80\% of the blended flux selects relatively simple events, but excludes potentially
informative events in more crowded environments. The restriction of the event peaking within the GBTDS also excludes some events whose astrometric signatures could be measurable during the survey. Relaxing these detection criteria could lead to a larger yield of BH events.

Finally, our single-lens calculations do not assess the effects of lens or source multiplicity. Binary motion can introduce additional astrometric structure, alter the lensing signal, and
complicate lens classification \citep[e.g.,][]{Abrams2025binary}. Quantifying these effects could be particularly important for events without significant photometric microlensing signals.

\subsection{Implications for the GBTDS}

The astrometric timescale distribution peaks near 1000 days, and most detectable BH signals extend across multiple high-cadence observing seasons. Searches restricted to individual seasons will therefore be poorly matched to much of this BH population. Computation of residuals and summary statistics such as UWE from standard astrometric source fits would
enable searches for microlensing BHs that complement the MSOS photometric microlensing pipeline. As the timescales of astrometric signals caused by BH lenses are comparable to the nominal \textit{Roman} mission duration, the long gaps between observing seasons hinder event characterization. Improving the precision of low-cadence gap-filling GBTDS observations (e.g. by enabling stacking of exposures) could therefore improve constraints on event parameters.

Observations before or after the nominal GBTDS would provide a complementary benefit by extending the astrometric baseline \citep[e.g., pre-GBTDS \textit{Gaia} data;][]{Prusti2016}. Precursor observations  \citep[e.g.,][]{Euclid2026}, an extended \textit{Roman} mission, or targeted follow-up could help distinguish a long-duration microlensing deflection from the source's unlensed motion.

In summary, astrometry from the Roman GBTDS can provide an independent discovery channel for isolated BHs, rather than serving only to characterize photometrically selected events. Our simulations predict several hundred detectable BH signals, approximately four-fifths of which will be difficult to detect photometrically. Precise mass measurements will remain challenging for many of these events, but joint photometric and astrometric modelling and probabilistic lens classification could turn this sample into a valuable probe of isolated-BH abundance. Realizing this opportunity will require dedicated multi-season astrometric variability searches, astrometric selection functions, and careful treatment of competing sources of astrometric variability.

\section*{Acknowledgements}
\noindent This work was performed under the auspices of the U.S. Department of Energy by Lawrence Livermore National Laboratory under Contract DE-AC52-07NA27344. The document number is LLNL-JRNL-2026695. This work was supported by the LLNL-LDRD Program under Project 27-ERD-017. The authors acknowledge support by the state of Baden-Württemberg through bwHPC.

\bibliography{refs}{}

@ARTICLE{Abrams2025binary,
       author = {{Abrams}, Natasha S. and {Lu}, Jessica R. and {Lam}, Casey Y. and {Medford}, Michael S. and {Hosek}, Matthew W. and {Rose}, Sam},
        title = "{Assessing the Impact of Binary Systems on Microlensing Using SPISEA and PopSyCLE Population Simulations}",
      journal = {\apj},
         year = 2025,
        month = feb,
       volume = {980},
       number = {1},
          eid = {103},
        pages = {103},
          doi = {10.3847/1538-4357/ada5f9},
archivePrefix = {arXiv},
       eprint = {2501.03506},
 primaryClass = {astro-ph.SR},
       adsurl = {https://ui.adsabs.harvard.edu/abs/2025ApJ...980..103A}
}

@ARTICLE{Gould2000,
       author = {{Gould}, Andrew},
        title = "{A Natural Formalism for Microlensing}",
      journal = {\apj},
         year = 2000,
        month = oct,
       volume = {542},
       number = {2},
        pages = {785-788},
          doi = {10.1086/317037},
archivePrefix = {arXiv},
       eprint = {astro-ph/0001421},
 primaryClass = {astro-ph},
       adsurl = {https://ui.adsabs.harvard.edu/abs/2000ApJ...542..785G}
}

@ARTICLE{Higson2019,
       author = {{Higson}, Edward and {Handley}, Will and {Hobson}, Mike and {Lasenby}, Anthony},
        title = "{Dynamic nested sampling: an improved algorithm for parameter estimation and evidence calculation}",
      journal = {Statistics and Computing},
         year = 2019,
        month = sep,
       volume = {29},
       number = {5},
        pages = {891-913},
          doi = {10.1007/s11222-018-9844-0},
archivePrefix = {arXiv},
       eprint = {1704.03459},
 primaryClass = {stat.CO},
       adsurl = {https://ui.adsabs.harvard.edu/abs/2019S&C....29..891H}
}

@ARTICLE{Tilburg2018,
       author = {{Van Tilburg}, Ken and {Taki}, Anna-Maria and {Weiner}, Neal},
        title = "{Halometry from astrometry}",
      journal = {\jcap},
         year = 2018,
        month = jul,
       volume = {2018},
       number = {7},
          eid = {041},
        pages = {041},
          doi = {10.1088/1475-7516/2018/07/041},
archivePrefix = {arXiv},
       eprint = {1804.01991},
 primaryClass = {astro-ph.CO},
       adsurl = {https://ui.adsabs.harvard.edu/abs/2018JCAP...07..041V}
}

@ARTICLE{Verma2023,
       author = {{Verma}, Himanshu and {Rentala}, Vikram},
        title = "{Astrometric microlensing of primordial black holes with Gaia}",
      journal = {\jcap},
         year = 2023,
        month = may,
       volume = {2023},
       number = {5},
          eid = {045},
        pages = {045},
          doi = {10.1088/1475-7516/2023/05/045},
archivePrefix = {arXiv},
       eprint = {2208.14460},
 primaryClass = {astro-ph.GA},
       adsurl = {https://ui.adsabs.harvard.edu/abs/2023JCAP...05..045V}
}

@ARTICLE{Perkins2026,
       author = {{Perkins}, Scott E. and {McGill}, Peter and {Dawson}, William A. and {Ho}, Ming-Feng and {Abrams}, Natasha S. and {Bird}, Simeon and {Lu}, Jessica R.},
        title = "{Microlensing Constraints on Primordial Black Holes toward the Bulge Are Galactic Model Limited}",
      journal = {\apj},
         year = 2026,
        month = jul,
       volume = {1006},
       number = {1},
          eid = {110},
        pages = {110},
          doi = {10.3847/1538-4357/ae66f6},
       adsurl = {https://ui.adsabs.harvard.edu/abs/2026ApJ..1006..110P}
}

@ARTICLE{Lam2026,
       author = {{Lam}, Casey Y. and {Lu}, Jessica R. and {El-Badry}, Kareem and {Seth}, Anil and {Simon}, Joshua D. and {Abrams}, Natasha S. and {Bellini}, Andrea and {Hosek}, Jr., Matthew W. and {McGill}, Peter},
        title = "{Black hole astrometric binaries in the Roman Galactic Bulge Time Domain Survey}",
      journal = {arXiv e-prints},
         year = 2026,
        month = aug,
          eid = {arXiv:2608.24998},
        pages = {arXiv:2608.24998},
          doi = {10.48550/arXiv.2608.24998},
archivePrefix = {arXiv},
       eprint = {2608.24998},
 primaryClass = {astro-ph.SR},
       adsurl = {https://ui.adsabs.harvard.edu/abs/2026arXiv260824998L}
}

@ARTICLE{Sahu2017,
       author = {{Sahu}, Kailash C. and {Anderson}, Jay and {Casertano}, Stefano and {Bond}, Howard E. and {Bergeron}, Pierre and {Nelan}, Edmund P. and {Pueyo}, Laurent and {Brown}, Thomas M. and {Bellini}, Andrea and {Levay}, Zoltan G. and {Sokol}, Joshua and {Dominik}, Martin and {Calamida}, Annalisa and {Kains}, No{\'e} and {Livio}, Mario},
        title = "{Relativistic deflection of background starlight measures the mass of a nearby white dwarf star}",
      journal = {Science},
         year = 2017,
        month = jun,
       volume = {356},
       number = {6342},
        pages = {1046-1050},
          doi = {10.1126/science.aal2879},
archivePrefix = {arXiv},
       eprint = {1706.02037},
 primaryClass = {astro-ph.SR},
       adsurl = {https://ui.adsabs.harvard.edu/abs/2017Sci...356.1046S}
}

@ARTICLE{Howil2025,
       author = {{Howil}, K. and {Wyrzykowski}, {\L}. and {Kruszy{\'n}ska}, K. and {Zieli{\'n}ski}, P. and {Bachelet}, E. and {Gromadzki}, M. and {Miko{\l}ajczyk}, P.~J. and {Kotysz}, K. and {Jab{\l}o{\'n}ska}, M. and {Kaczmarek}, Z. and {Mr{\'o}z}, P. and {Ihanec}, N. and {Ratajczak}, M. and {Pylypenko}, U. and {Rybicki}, K. and {Sweeney}, D. and {Hodgkin}, S.~T. and {Larma}, M. and {Carrasco}, J.~M. and {Burgaz}, U. and {Godunova}, V. and {Simon}, A. and {Cusano}, F. and {Jelinek}, M. and {{\v{S}}trobl}, J. and {Hudec}, R. and {Merc}, J. and {Ku{\v{c}}{\'a}kov{\'a}}, H. and {Erece}, O. and {Kilic}, Y. and {Olivares}, F. and {Morrell}, M. and {Wicker}, M.},
        title = "{Uncovering the invisible: A study of Gaia18ajz, a candidate black hole revealed by microlensing}",
      journal = {\aap},
         year = 2025,
        month = feb,
       volume = {694},
          eid = {A94},
        pages = {A94},
          doi = {10.1051/0004-6361/202451046},
archivePrefix = {arXiv},
       eprint = {2403.09006},
 primaryClass = {astro-ph.GA},
       adsurl = {https://ui.adsabs.harvard.edu/abs/2025A&A...694A..94H}
}

@ARTICLE{Ozel2010,
       author = {{{\"O}zel}, Feryal and {Psaltis}, Dimitrios and {Narayan}, Ramesh and {McClintock}, Jeffrey E.},
        title = "{The Black Hole Mass Distribution in the Galaxy}",
      journal = {\apj},
         year = 2010,
        month = dec,
       volume = {725},
       number = {2},
        pages = {1918-1927},
          doi = {10.1088/0004-637X/725/2/1918},
archivePrefix = {arXiv},
       eprint = {1006.2834},
 primaryClass = {astro-ph.GA},
       adsurl = {https://ui.adsabs.harvard.edu/abs/2010ApJ...725.1918O}
}

@ARTICLE{Fryer2012,
       author = {{Fryer}, Chris L. and {Belczynski}, Krzysztof and {Wiktorowicz}, Grzegorz and {Dominik}, Michal and {Kalogera}, Vicky and {Holz}, Daniel E.},
        title = "{Compact Remnant Mass Function: Dependence on the Explosion Mechanism and Metallicity}",
      journal = {\apj},
         year = 2012,
        month = apr,
       volume = {749},
       number = {1},
          eid = {91},
        pages = {91},
          doi = {10.1088/0004-637X/749/1/91},
archivePrefix = {arXiv},
       eprint = {1110.1726},
 primaryClass = {astro-ph.SR},
       adsurl = {https://ui.adsabs.harvard.edu/abs/2012ApJ...749...91F}
}

@ARTICLE{Safizadeh1999,
       author = {{Safizadeh}, Neda and {Dalal}, Neal and {Griest}, Kim},
        title = "{Astrometric Microlensing as a Method of Discovering and Characterizing Extrasolar Planets}",
      journal = {\apj},
         year = 1999,
        month = sep,
       volume = {522},
       number = {1},
        pages = {512-517},
          doi = {10.1086/307628},
archivePrefix = {arXiv},
       eprint = {astro-ph/9811233},
 primaryClass = {astro-ph},
       adsurl = {https://ui.adsabs.harvard.edu/abs/1999ApJ...522..512S}
}

@article{Olejak2022,
   title={The role of supernova convection for the lower mass gap in the isolated binary formation of gravitational wave sources},
   volume={516},
   ISSN={1365-2966},
   url={http://dx.doi.org/10.1093/mnras/stac2359},
   DOI={10.1093/mnras/stac2359},
   number={2},
   journal={Monthly Notices of the Royal Astronomical Society},
   publisher={Oxford University Press (OUP)},
   author={Olejak, Aleksandra and Fryer, Chris L and Belczynski, Krzysztof and Baibhav, Vishal},
   year={2022},
   month=aug, pages={2252–2271} }

@ARTICLE{Bailyn1998,
       author = {{Bailyn}, Charles D. and {Jain}, Raj K. and {Coppi}, Paolo and {Orosz}, Jerome A.},
        title = "{The Mass Distribution of Stellar Black Holes}",
      journal = {\apj},
         year = 1998,
        month = may,
       volume = {499},
       number = {1},
        pages = {367-374},
          doi = {10.1086/305614},
archivePrefix = {arXiv},
       eprint = {astro-ph/9708032},
 primaryClass = {astro-ph},
       adsurl = {https://ui.adsabs.harvard.edu/abs/1998ApJ...499..367B}
}

@misc{RomanMSOSForROSES,
  author = {{Roman Science Support Center}},
  year = 2024,
  title        = {Microlensing Science Operations System (MSOS) for ROSES},
  howpublished = {\url{https://roman.gsfc.nasa.gov/science/roses/MSOSForROSES.pdf}},
  note         = {Accessed: 2026-04-16},
  
}

@ARTICLE{Lindegren2021,
       author = {{Lindegren}, L. and {Klioner}, S.~A. and {Hern{\'a}ndez}, J. and {Bombrun}, A. and {Ramos-Lerate}, M. and {Steidelm{\"u}ller}, H. and {Bastian}, U. and {Biermann}, M. and {de Torres}, A. and {Gerlach}, E. and {Geyer}, R. and {Hilger}, T. and {Hobbs}, D. and {Lammers}, U. and {McMillan}, P.~J. and {Stephenson}, C.~A. and {Casta{\~n}eda}, J. and {Davidson}, M. and {Fabricius}, C. and {Gracia-Abril}, G. and {Portell}, J. and {Rowell}, N. and {Teyssier}, D. and {Torra}, F. and {Bartolom{\'e}}, S. and {Clotet}, M. and {Garralda}, N. and {Gonz{\'a}lez-Vidal}, J.~J. and {Torra}, J. and {Abbas}, U. and {Altmann}, M. and {Anglada Varela}, E. and {Balaguer-N{\'u}{\~n}ez}, L. and {Balog}, Z. and {Barache}, C. and {Becciani}, U. and {Bernet}, M. and {Bertone}, S. and {Bianchi}, L. and {Bouquillon}, S. and {Brown}, A.~G.~A. and {Bucciarelli}, B. and {Busonero}, D. and {Butkevich}, A.~G. and {Buzzi}, R. and {Cancelliere}, R. and {Carlucci}, T. and {Charlot}, P. and {Cioni}, M. -R.~L. and {Crosta}, M. and {Crowley}, C. and {del Peloso}, E.~F. and {del Pozo}, E. and {Drimmel}, R. and {Esquej}, P. and {Fienga}, A. and {Fraile}, E. and {Gai}, M. and {Garcia-Reinaldos}, M. and {Guerra}, R. and {Hambly}, N.~C. and {Hauser}, M. and {Jan{\ss}en}, K. and {Jordan}, S. and {Kostrzewa-Rutkowska}, Z. and {Lattanzi}, M.~G. and {Liao}, S. and {Licata}, E. and {Lister}, T.~A. and {L{\"o}ffler}, W. and {Marchant}, J.~M. and {Masip}, A. and {Mignard}, F. and {Mints}, A. and {Molina}, D. and {Mora}, A. and {Morbidelli}, R. and {Murphy}, C.~P. and {Pagani}, C. and {Panuzzo}, P. and {Pe{\~n}alosa Esteller}, X. and {Poggio}, E. and {Re Fiorentin}, P. and {Riva}, A. and {Sagrist{\`a} Sell{\'e}s}, A. and {Sanchez Gimenez}, V. and {Sarasso}, M. and {Sciacca}, E. and {Siddiqui}, H.~I. and {Smart}, R.~L. and {Souami}, D. and {Spagna}, A. and {Steele}, I.~A. and {Taris}, F. and {Utrilla}, E. and {van Reeven}, W. and {Vecchiato}, A.},
        title = "{Gaia Early Data Release 3. The astrometric solution}",
      journal = {\aap},
         year = 2021,
        month = may,
       volume = {649},
          eid = {A2},
        pages = {A2},
          doi = {10.1051/0004-6361/202039709},
archivePrefix = {arXiv},
       eprint = {2012.03380},
 primaryClass = {astro-ph.IM},
       adsurl = {https://ui.adsabs.harvard.edu/abs/2021A&A...649A...2L}
}

@article{ROTAC,
  author = {Zasowski, Gail and Jha, Saurabh W. and others},
  title = {Roman Observations Time Allocation Committee: Final Report and Recommendations},
  year = {2025},
  eprint = {2505.10574},
  archivePrefix = {arXiv},
  doi = {10.48550/arXiv.2505.10574}
}

@ARTICLE{Farr2011,
       author = {{Farr}, Will M. and {Sravan}, Niharika and {Cantrell}, Andrew and {Kreidberg}, Laura and {Bailyn}, Charles D. and {Mandel}, Ilya and {Kalogera}, Vicky},
        title = "{The Mass Distribution of Stellar-mass Black Holes}",
      journal = {\apj},
         year = 2011,
        month = nov,
       volume = {741},
       number = {2},
          eid = {103},
        pages = {103},
          doi = {10.1088/0004-637X/741/2/103},
archivePrefix = {arXiv},
       eprint = {1011.1459},
 primaryClass = {astro-ph.GA},
       adsurl = {https://ui.adsabs.harvard.edu/abs/2011ApJ...741..103F}
}

@ARTICLE{Belczynski2012,
       author = {{Belczynski}, Krzysztof and {Wiktorowicz}, Grzegorz and {Fryer}, Chris L. and {Holz}, Daniel E. and {Kalogera}, Vassiliki},
        title = "{Missing Black Holes Unveil the Supernova Explosion Mechanism}",
      journal = {\apj},
         year = 2012,
        month = sep,
       volume = {757},
       number = {1},
          eid = {91},
        pages = {91},
          doi = {10.1088/0004-637X/757/1/91},
archivePrefix = {arXiv},
       eprint = {1110.1635},
 primaryClass = {astro-ph.GA},
       adsurl = {https://ui.adsabs.harvard.edu/abs/2012ApJ...757...91B}
}

@ARTICLE{Kruszynska2024,
       author = {{Kruszy{\'n}ska}, K. and {Wyrzykowski}, {\L}. and {Rybicki}, K.~A. and {Howil}, K. and {Jab{\l}o{\'n}ska}, M. and {Kaczmarek}, Z. and {Ihanec}, N. and {Maskoli{\={u}}nas}, M. and {Bronikowski}, M. and {Pylypenko}, U. and {Udalski}, A. and {Mr{\'o}z}, P. and {Poleski}, R. and {Skowron}, J. and {Szyma{\'n}ski}, M.~K. and {Soszy{\'n}ski}, I. and {Pietrukowicz}, P. and {Koz{\l}owski}, S. and {Ulaczyk}, K. and {Iwanek}, P. and {Wrona}, M. and {Gromadzki}, M. and {Mr{\'o}z}, M.~J. and {Abe}, F. and {Bando}, K. and {Barry}, R. and {Bennett}, D.~P. and {Bhattacharya}, A. and {Bond}, I.~A. and {Fukui}, A. and {Hamada}, R. and {Hamada}, S. and {Hamasaki}, N. and {Hirao}, Y. and {Ishitani Silva}, S. and {Itow}, Y. and {Koshimoto}, N. and {Matsubara}, Y. and {Miyazaki}, S. and {Muraki}, Y. and {Nagai}, T. and {Nunota}, K. and {Olmschenk}, G. and {Ranc}, C. and {Rattenbury}, N.~J. and {Satoh}, Y. and {Sumi}, T. and {Suzuki}, D. and {Tristram}, P.~J. and {Vandorou}, A. and {Yama}, H.},
        title = "{Dark lens candidates from Gaia Data Release 3}",
      journal = {\aap},
         year = 2024,
        month = dec,
       volume = {692},
          eid = {A28},
        pages = {A28},
          doi = {10.1051/0004-6361/202449322},
archivePrefix = {arXiv},
       eprint = {2401.13759},
 primaryClass = {astro-ph.GA},
       adsurl = {https://ui.adsabs.harvard.edu/abs/2024A&A...692A..28K}
}

@article{Spearman1904,
 ISSN = {00029556},
 URL = {http://www.jstor.org/stable/1412159},
 author = {C. Spearman},
 journal = {The American Journal of Psychology},
 number = {1},
 pages = {72--101},
 publisher = {University of Illinois Press},
 title = {The Proof and Measurement of Association between Two Things},
 urldate = {2026-11-10},
 volume = {15},
 year = {1904},
 doi={10.2307/1412159}
}

@ARTICLE{GaiaEDR3val,
       author = {{Fabricius}, C. and {Luri}, X. and {Arenou}, F. and {Babusiaux}, C. and {Helmi}, A. and {Muraveva}, T. and {Reyl{\'e}}, C. and {Spoto}, F. and {Vallenari}, A. and {Antoja}, T. and {Balbinot}, E. and {Barache}, C. and {Bauchet}, N. and {Bragaglia}, A. and {Busonero}, D. and {Cantat-Gaudin}, T. and {Carrasco}, J.~M. and {Diakit{\'e}}, S. and {Fabrizio}, M. and {Figueras}, F. and {Garcia-Gutierrez}, A. and {Garofalo}, A. and {Jordi}, C. and {Kervella}, P. and {Khanna}, S. and {Leclerc}, N. and {Licata}, E. and {Lambert}, S. and {Marrese}, P.~M. and {Masip}, A. and {Ramos}, P. and {Robichon}, N. and {Robin}, A.~C. and {Romero-G{\'o}mez}, M. and {Rubele}, S. and {Weiler}, M.},
        title = "{Gaia Early Data Release 3. Catalogue validation}",
      journal = {\aap},
         year = 2021,
        month = may,
       volume = {649},
          eid = {A5},
        pages = {A5},
          doi = {10.1051/0004-6361/202039834},
archivePrefix = {arXiv},
       eprint = {2012.06242},
 primaryClass = {astro-ph.GA},
       adsurl = {https://ui.adsabs.harvard.edu/abs/2021A&A...649A...5F}
}

@ARTICLE{Holl2023,
       author = {{Holl}, B. and {Fabricius}, C. and {Portell}, J. and {Lindegren}, L. and {Panuzzo}, P. and {Bernet}, M. and {Casta{\~n}eda}, J. and {Jevardat de Fombelle}, G. and {Audard}, M. and {Ducourant}, C. and {Harrison}, D.~L. and {Evans}, D.~W. and {Busso}, G. and {Sozzetti}, A. and {Gosset}, E. and {Arenou}, F. and {De Angeli}, F. and {Riello}, M. and {Eyer}, L. and {Rimoldini}, L. and {Gavras}, P. and {Mowlavi}, N. and {Nienartowicz}, K. and {Lecoeur-Ta{\"\i}bi}, I. and {Garc{\'\i}a-Lario}, P. and {Pourbaix}, D.},
        title = "{Gaia Data Release 3. Gaia scan-angle-dependent signals and spurious periods}",
      journal = {\aap},
         year = 2023,
        month = jun,
       volume = {674},
          eid = {A25},
        pages = {A25},
          doi = {10.1051/0004-6361/202245353},
archivePrefix = {arXiv},
       eprint = {2212.11971},
 primaryClass = {astro-ph.IM},
       adsurl = {https://ui.adsabs.harvard.edu/abs/2023A&A...674A..25H}
}

@misc{RDox,
    author       = {{Roman User Documentation}},
    title        = "{Roman User Documentation (RDox)}",
    year         = {2022--},
    publisher    = {Space Telescope Science Institute and Caltech/IPAC},
    address      = {Baltimore, MD and Pasadena, CA},
    url          = {https://roman-docs.stsci.edu/}
}

@article{Jablonska2022,
	author = {{Jabłońska}, Maja and {Wyrzykowski}, \Lukasz and {Rybicki}, Krzysztof A. and {Kruszyńska, Katarzyna} and {Kaczmarek, Zofia} and {Penoyre, Zephyr}},
	title = {A possible nearby microlensing stellar remnant hiding in Gaia DR3 astrometry},
	DOI= "10.1051/0004-6361/202244656",
	url= "https://doi.org/10.1051/0004-6361/202244656",
	journal = {A\&A},
	year = 2022,
	volume = 666,
	pages = "L16",
}

@ARTICLE{Wyrzykowski2016,
       author = {{Wyrzykowski}, {\L}. and {Kostrzewa-Rutkowska}, Z. and {Skowron}, J. and {Rybicki}, K.~A. and {Mr{\'o}z}, P. and {Koz{\l}owski}, S. and {Udalski}, A. and {Szyma{\'n}ski}, M.~K. and {Pietrzy{\'n}ski}, G. and {Soszy{\'n}ski}, I. and {Ulaczyk}, K. and {Pietrukowicz}, P. and {Poleski}, R. and {Pawlak}, M. and {I{\l}kiewicz}, K. and {Rattenbury}, N.~J.},
        title = "{Black hole, neutron star and white dwarf candidates from microlensing with OGLE-III}",
      journal = {\mnras},
         year = 2016,
        month = may,
       volume = {458},
       number = {3},
        pages = {3012-3026},
          doi = {10.1093/mnras/stw426},
archivePrefix = {arXiv},
       eprint = {1509.04899},
 primaryClass = {astro-ph.SR},
       adsurl = {https://ui.adsabs.harvard.edu/abs/2016MNRAS.458.3012W}
}

@ARTICLE{Kreidberg2012,
       author = {{Kreidberg}, Laura and {Bailyn}, Charles D. and {Farr}, Will M. and {Kalogera}, Vicky},
        title = "{Mass Measurements of Black Holes in X-Ray Transients: Is There a Mass Gap?}",
      journal = {\apj},
         year = 2012,
        month = sep,
       volume = {757},
       number = {1},
          eid = {36},
        pages = {36},
          doi = {10.1088/0004-637X/757/1/36},
archivePrefix = {arXiv},
       eprint = {1205.1805},
 primaryClass = {astro-ph.HE},
       adsurl = {https://ui.adsabs.harvard.edu/abs/2012ApJ...757...36K}
}

@ARTICLE{Rybicki2024,
       author = {{Rybicki}, Krzysztof A. and {Shvartzvald}, Yossi and {Yee}, Jennifer C. and {Calchi Novati}, Sebastiano and {Ofek}, Eran O. and {Bond}, Ian A. and {Beichman}, Charles and {Bryden}, Geoff and {Carey}, Sean and {Henderson}, Calen and {Zhu}, Wei and {Fausnaugh}, Michael M. and {Wibking}, Benjamin and {Udalski}, Andrzej and {Poleski}, Radek and {Mr{\'o}z}, Przemek and {Szyma{\'n}ski}, Micha{\l} K. and {Soszy{\'n}ski}, Igor and {Pietrukowicz}, Pawe{\l} and {Koz{\l}owski}, Szymon and {Skowron}, Jan and {Ulaczyk}, Krzysztof and {Iwanek}, Patryk and {Wrona}, Marcin and {Ryu}, Yoon-Hyun and {Albrow}, Michael D. and {Chung}, Sun-Ju and {Gould}, Andrew and {Han}, Cheongho and {Hwang}, Kyu-Ha and {Jung}, Youn Kil and {Shin}, In-Gu and {Yang}, Hongjing and {Zang}, Weicheng and {Cha}, Sang-Mok and {Kim}, Dong-Jin and {Kim}, Hyoun-Woo and {Kim}, Seung-Lee and {Lee}, Chung-Uk and {Lee}, Dong-Joo and {Lee}, Yongseok and {Park}, Byeong-Gon and {Pogge}, Richard W. and {Abe}, Fumio and {Barry}, Richard and {Bennett}, David P. and {Bhattacharya}, Aparna and {Fukui}, Akihiko and {Hamada}, Ryusei and {Hamada}, Shunya and {Hamasaki}, Naoto and {Hirao}, Yuki and {Ishitani Silva}, Stela and {Itow}, Yoshitaka and {Kirikawa}, Rintaro and {Koshimoto}, Naoki and {Matsubara}, Yutaka and {Miyazaki}, Shota and {Muraki}, Yasushi and {Nagai}, Tutumi and {Nunota}, Kansuke and {Olmschenk}, Greg and {Ranc}, Clement and {Rattenbury}, Nicholas J. and {Satoh}, Yuki K. and {Sumi}, Takahiro and {Suzuki}, Daisuke and {Tristram}, Paul. J. and {Vandorou}, Aikaterini and {Yama}, Hibiki and {Wyrzykowski}, {\L}ukasz and {Howil}, Kornel and {Kruszy{\'n}ska}, Katarzyna},
        title = "{Analysis of the Full Spitzer Microlensing Sample. I. Dark Remnant Candidates and Gaia Predictions}",
      journal = {\apj},
         year = 2024,
        month = nov,
       volume = {975},
       number = {2},
          eid = {216},
        pages = {216},
          doi = {10.3847/1538-4357/ad7bb1},
archivePrefix = {arXiv},
       eprint = {2407.13740},
 primaryClass = {astro-ph.GA},
       adsurl = {https://ui.adsabs.harvard.edu/abs/2024ApJ...975..216R}
}

@ARTICLE{Sallaberry2025,
       author = {{Sallaberry}, Greg and {Kaczmarek}, Zofia and {McGill}, Peter and {Perkins}, Scott and {Dawson}, William and {Begbie}, Caitlin},
        title = "{popclass: A Python Package for Classifying Microlensing Events}",
      journal = {The Journal of Open Source Software},
         year = 2025,
        month = may,
       volume = {10},
       number = {109},
          eid = {7769},
        pages = {7769},
          doi = {10.21105/joss.07769},
archivePrefix = {arXiv},
       eprint = {2410.14076},
 primaryClass = {astro-ph.IM},
       adsurl = {https://ui.adsabs.harvard.edu/abs/2025JOSS...10.7769S}
}

@ARTICLE{Samland1998,
       author = {{Samland}, M.},
        title = "{Modeling the Evolution of Disk Galaxies. II. Yields of Massive Stars}",
      journal = {\apj},
         year = 1998,
        month = mar,
       volume = {496},
       number = {1},
        pages = {155-171},
          doi = {10.1086/305368},
       adsurl = {https://ui.adsabs.harvard.edu/abs/1998ApJ...496..155S}
}

@ARTICLE{LamandLu2023,
       author = {{Lam}, Casey Y. and {Lu}, Jessica R.},
        title = "{A Reanalysis of the Isolated Black Hole Candidate OGLE-2011-BLG-0462/MOA-2011-BLG-191}",
      journal = {\apj},
         year = 2023,
        month = oct,
       volume = {955},
       number = {2},
          eid = {116},
        pages = {116},
          doi = {10.3847/1538-4357/aced4a},
archivePrefix = {arXiv},
       eprint = {2308.03302},
 primaryClass = {astro-ph.SR},
       adsurl = {https://ui.adsabs.harvard.edu/abs/2023ApJ...955..116L}
}

@ARTICLE{Repetto2017,
       author = {{Repetto}, Serena and {Igoshev}, Andrei P. and {Nelemans}, Gijs},
        title = "{The Galactic distribution of X-ray binaries and its implications for compact object formation and natal kicks}",
      journal = {\mnras},
         year = 2017,
        month = may,
       volume = {467},
       number = {1},
        pages = {298-310},
          doi = {10.1093/mnras/stx027},
archivePrefix = {arXiv},
       eprint = {1701.01347},
 primaryClass = {astro-ph.HE},
       adsurl = {https://ui.adsabs.harvard.edu/abs/2017MNRAS.467..298R}
}

@ARTICLE{Ivanova2013,
       author = {{Ivanova}, N. and {Justham}, S. and {Chen}, X. and {De Marco}, O. and {Fryer}, C.~L. and {Gaburov}, E. and {Ge}, H. and {Glebbeek}, E. and {Han}, Z. and {Li}, X. -D. and {Lu}, G. and {Marsh}, T. and {Podsiadlowski}, P. and {Potter}, A. and {Soker}, N. and {Taam}, R. and {Tauris}, T.~M. and {van den Heuvel}, E.~P.~J. and {Webbink}, R.~F.},
        title = "{Common envelope evolution: where we stand and how we can move forward}",
      journal = {\aapr},
         year = 2013,
        month = feb,
       volume = {21},
          eid = {59},
        pages = {59},
          doi = {10.1007/s00159-013-0059-2},
archivePrefix = {arXiv},
       eprint = {1209.4302},
 primaryClass = {astro-ph.HE},
       adsurl = {https://ui.adsabs.harvard.edu/abs/2013A&ARv..21...59I}
}

@article{Ho2024,
  title = {Investigating the mixing between two black hole populations in LIGO-Virgo-KAGRA GWTC-3},
  author = {Ho, Ming-Feng and Perkins, Scott E. and Bird, Simeon and Dawson, William A. and Golovich, Nathan and Lu, Jessica R. and McGill, Peter},
  journal = {Phys. Rev. D},
  volume = {110},
  issue = {6},
  pages = {063031},
  numpages = {18},
  year = {2024},
  month = {Sep},
  publisher = {American Physical Society},
  doi = {10.1103/PhysRevD.110.063031},
  url = {https://link.aps.org/doi/10.1103/PhysRevD.110.063031}
}

@ARTICLE{Abbott2023,
       author = {{Abbott}, R. and {Abbott}, T.~D. and {Acernese}, F. and {Ackley}, K. and {Adams}, C. and {Adhikari}, N. and {Adhikari}, R.~X. and {Adya}, V.~B. and {Affeldt}, C. and {Agarwal}, D. and {Agathos}, M. and {Agatsuma}, K. and {Aggarwal}, N. and {Aguiar}, O.~D. and {Aiello}, L. and {Ain}, A. and {Ajith}, P. and {Akcay}, S. and {Akutsu}, T. and {Albanesi}, S. and {Allocca}, A. and {Altin}, P.~A. and {Amato}, A. and {Anand}, C. and {Anand}, S. and {Ananyeva}, A. and {Anderson}, S.~B. and {Anderson}, W.~G. and {Ando}, M. and {Andrade}, T. and {Andres}, N. and {Andri{\'c}}, T. and {Angelova}, S.~V. and {Ansoldi}, S. and {Antelis}, J.~M. and {Antier}, S. and {Appert}, S. and {Arai}, Koji and {Arai}, Koya and {Arai}, Y. and {Araki}, S. and {Araya}, A. and {Araya}, M.~C. and {Areeda}, J.~S. and {Ar{\`e}ne}, M. and {Aritomi}, N. and {Arnaud}, N. and {Arogeti}, M. and {Aronson}, S.~M. and {Arun}, K.~G. and {Asada}, H. and {Asali}, Y. and {Ashton}, G. and {Aso}, Y. and {Assiduo}, M. and {Aston}, S.~M. and {Astone}, P. and {Aubin}, F. and {Austin}, C. and {Babak}, S. and {Badaracco}, F. and {Bader}, M.~K.~M. and {Badger}, C. and {Bae}, S. and {Bae}, Y. and {Baer}, A.~M. and {Bagnasco}, S. and {Bai}, Y. and {Baiotti}, L. and {Baird}, J. and {Bajpai}, R. and {Ball}, M. and {Ballardin}, G. and {Ballmer}, S.~W. and {Balsamo}, A. and {Baltus}, G. and {Banagiri}, S. and {Bankar}, D. and {Barayoga}, J.~C. and {Barbieri}, C. and {Barish}, B.~C. and {Barker}, D. and {Barneo}, P. and {Barone}, F. and {Barr}, B. and {Barsotti}, L. and {Barsuglia}, M. and {Barta}, D. and {Bartlett}, J. and {Barton}, M.~A. and {Bartos}, I. and {Bassiri}, R. and {Basti}, A. and {Bawaj}, M. and {Bayley}, J.~C. and {Baylor}, A.~C. and {Bazzan}, M. and {B{\'e}csy}, B. and {Bedakihale}, V.~M. and {Bejger}, M. and {Belahcene}, I. and {Benedetto}, V. and {Beniwal}, D. and {Bennett}, T.~F. and {Bentley}, J.~D. and {Benyaala}, M. and {Bergamin}, F. and {Berger}, B.~K. and {Bernuzzi}, S. and {Berry}, C.~P.~L. and {Bersanetti}, D. and {Bertolini}, A. and {Betzwieser}, J. and {Beveridge}, D. and {Bhandare}, R. and {Bhardwaj}, U. and {Bhattacharjee}, D. and {Bhaumik}, S. and {Bilenko}, I.~A. and {Billingsley}, G. and {Bini}, S. and {Birney}, R. and {Birnholtz}, O. and {Biscans}, S. and {Bischi}, M. and {Biscoveanu}, S. and {Bisht}, A. and {Biswas}, B. and {Bitossi}, M. and {Bizouard}, M. -A. and {Blackburn}, J.~K. and {Blair}, C.~D. and {Blair}, D.~G. and {Blair}, R.~M. and {Bobba}, F. and {Bode}, N. and {Boer}, M. and {Bogaert}, G. and {Boldrini}, M. and {Bonavena}, L.~D. and {Bondu}, F. and {Bonilla}, E. and {Bonnand}, R. and {Booker}, P. and {Boom}, B.~A. and {Bork}, R. and {Boschi}, V. and {Bose}, N. and {Bose}, S. and {Bossilkov}, V. and {Boudart}, V. and {Bouffanais}, Y. and {Bozzi}, A. and {Bradaschia}, C. and {Brady}, P.~R. and {Bramley}, A. and {Branch}, A. and {Branchesi}, M. and {Brandt}, J. and {Brau}, J.~E. and {Breschi}, M. and {Briant}, T. and {Briggs}, J.~H. and {Brillet}, A. and {Brinkmann}, M. and {Brockill}, P. and {Brooks}, A.~F. and {Brooks}, J. and {Brown}, D.~D. and {Brunett}, S. and {Bruno}, G. and {Bruntz}, R. and {Bryant}, J. and {Bulik}, T. and {Bulten}, H.~J. and {Buonanno}, A. and {Buscicchio}, R. and {Buskulic}, D. and {Buy}, C. and {Byer}, R.~L. and {Davies}, G.~S. Cabourn and {Cadonati}, L. and {Cagnoli}, G. and {Cahillane}, C. and {Bustillo}, J. Calder{\'o}n and {Callaghan}, J.~D. and {Callister}, T.~A. and {Calloni}, E. and {Cameron}, J. and {Camp}, J.~B. and {Canepa}, M. and {Canevarolo}, S. and {Cannavacciuolo}, M. and {Cannon}, K.~C. and {Cao}, H. and {Cao}, Z. and {Capocasa}, E. and {Capote}, E. and {Carapella}, G. and {Carbognani}, F. and {Carlin}, J.~B. and {Carney}, M.~F. and {Carpinelli}, M. and {Carrillo}, G. and {Carullo}, G. and {Carver}, T.~L. and {Diaz}, J. Casanueva and {Casentini}, C. and {Castaldi}, G. and {Caudill}, S. and {Cavagli{\`a}}, M. and {Cavalier}, F. and {Cavalieri}, R. and {Ceasar}, M. and {Cella}, G. and {Cerd{\'a}-Dur{\'a}n}, P. and {Cesarini}, E. and {Chaibi}, W. and {Chakravarti}, K. and {Subrahmanya}, S. Chalathadka and {Champion}, E. and {Chan}, C. -H. and {Chan}, C. and {Chan}, C.~L. and {Chan}, K. and {Chan}, M. and {Chandra}, K. and {Chanial}, P. and {Chao}, S. and {Chapman-Bird}, C.~E.~A. and {Charlton}, P. and {Chase}, E.~A. and {Chassande-Mottin}, E. and {Chatterjee}, C. and {Chatterjee}, Debarati and {Chatterjee}, Deep and {Chaturvedi}, M. and {Chaty}, S. and {Chatziioannou}, K. and {Chen}, C. and {Chen}, H.~Y. and {Chen}, J. and {Chen}, K. and {Chen}, X. and {Chen}, Y. -B. and {Chen}, Y. -R. and {Chen}, Z. and {Cheng}, H. and {Cheong}, C.~K. and {Cheung}, H.~Y. and {Chia}, H.~Y. and {Chiadini}, F. and {Chiang}, C. -Y. and {Chiarini}, G. and {Chierici}, R. and {Chincarini}, A. and {Chiofalo}, M.~L. and {Chiummo}, A. and {Cho}, G. and {Cho}, H.~S. and {Choudhary}, R.~K. and {Choudhary}, S. and {Christensen}, N. and {Chu}, H. and {Chu}, Q. and {Chu}, Y. -K. and {Chua}, S. and {Chung}, K.~W. and {Ciani}, G. and {Ciecielag}, P. and {Cie{\'s}lar}, M. and {Cifaldi}, M. and {Ciobanu}, A.~A. and {Ciolfi}, R. and {Cipriano}, F. and {Cirone}, A. and {Clara}, F. and {Clark}, E.~N. and {Clark}, J.~A. and {Clarke}, L. and {Clearwater}, P. and {Clesse}, S. and {Cleva}, F. and {Coccia}, E. and {Codazzo}, E. and {Cohadon}, P. -F. and {Cohen}, D.~E. and {Cohen}, L. and {Colleoni}, M. and {Collette}, C.~G. and {Colombo}, A. and {Colpi}, M. and {Compton}, C.~M. and {Constancio}, M. and {Conti}, L. and {Cooper}, S.~J. and {Corban}, P. and {Corbitt}, T.~R. and {Cordero-Carri{\'o}n}, I. and {Corezzi}, S. and {Corley}, K.~R. and {Cornish}, N. and {Corre}, D. and {Corsi}, A. and {Cortese}, S. and {Costa}, C.~A. and {Cotesta}, R. and {Coughlin}, M.~W. and {Coulon}, J. -P. and {Countryman}, S.~T. and {Cousins}, B. and {Couvares}, P. and {Coward}, D.~M. and {Cowart}, M.~J. and {Coyne}, D.~C. and {Coyne}, R. and {Creighton}, J.~D.~E. and {Creighton}, T.~D. and {Criswell}, A.~W. and {Croquette}, M. and {Crowder}, S.~G. and {Cudell}, J.~R. and {Cullen}, T.~J. and {Cumming}, A. and {Cummings}, R. and {Cunningham}, L. and {Cuoco}, E. and {Cury{\l}o}, M. and {Dabadie}, P. and {Canton}, T. Dal and {Dall'Osso}, S. and {D{\'a}lya}, G. and {Dana}, A. and {Daneshgaranbajastani}, L.~M. and {D'Angelo}, B. and {Danila}, B. and {Danilishin}, S. and {D'Antonio}, S. and {Danzmann}, K. and {Darsow-Fromm}, C. and {Dasgupta}, A. and {Datrier}, L.~E.~H. and {Dattilo}, V. and {Dave}, I. and {Davier}, M. and {Davis}, D. and {Davis}, M.~C. and {Daw}, E.~J. and {de Alarc{\'o}n}, P.~F. and {Dean}, R. and {Debra}, D. and {Deenadayalan}, M. and {Degallaix}, J. and {de Laurentis}, M. and {Del{\'e}glise}, S. and {Del Favero}, V. and {de Lillo}, F. and {de Lillo}, N. and {Del Pozzo}, W. and {Demarchi}, L.~M. and {de Matteis}, F. and {D'Emilio}, V. and {Demos}, N. and {Dent}, T. and {Depasse}, A. and {de Pietri}, R. and {De Rosa}, R. and {de Rossi}, C. and {Desalvo}, R. and {de Simone}, R. and {Dhurandhar}, S. and {D{\'\i}az}, M.~C. and {Diaz-Ortiz}, M. and {Didio}, N.~A. and {Dietrich}, T. and {di Fiore}, L. and {di Fronzo}, C. and {di Giorgio}, C. and {di Giovanni}, F. and {di Giovanni}, M. and {di Girolamo}, T. and {di Lieto}, A. and {Ding}, B. and {di Pace}, S. and {di Palma}, I. and {di Renzo}, F. and {Divakarla}, A.~K. and {Dmitriev}, A. and {Doctor}, Z. and {D'Onofrio}, L. and {Donovan}, F. and {Dooley}, K.~L. and {Doravari}, S. and {Dorrington}, I. and {Drago}, M. and {Driggers}, J.~C. and {Drori}, Y. and {Ducoin}, J. -G. and {Dupej}, P. and {Durante}, O. and {D'Urso}, D. and {Duverne}, P. -A. and {Dwyer}, S.~E. and {Eassa}, C. and {Easter}, P.~J. and {Ebersold}, M. and {Eckhardt}, T. and {Eddolls}, G. and {Edelman}, B. and {Edo}, T.~B. and {Edy}, O. and {Effler}, A. and {Eguchi}, S. and {Eichholz}, J. and {Eikenberry}, S.~S. and {Eisenmann}, M. and {Eisenstein}, R.~A. and {Ejlli}, A. and {Engelby}, E. and {Enomoto}, Y. and {Errico}, L. and {Essick}, R.~C. and {Estell{\'e}s}, H. and {Estevez}, D. and {Etienne}, Z. and {Etzel}, T. and {Evans}, M. and {Evans}, T.~M. and {Ewing}, B.~E. and {Fafone}, V. and {Fair}, H. and {Fairhurst}, S. and {Farah}, A.~M. and {Farinon}, S. and {Farr}, B. and {Farr}, W.~M. and {Farrow}, N.~W. and {Fauchon-Jones}, E.~J. and {Favaro}, G. and {Favata}, M. and {Fays}, M. and {Fazio}, M. and {Feicht}, J. and {Fejer}, M.~M. and {Fenyvesi}, E. and {Ferguson}, D.~L. and {Fernandez-Galiana}, A. and {Ferrante}, I. and {Ferreira}, T.~A. and {Fidecaro}, F. and {Figura}, P. and {Fiori}, I. and {Fishbach}, M. and {Fisher}, R.~P. and {Fittipaldi}, R. and {Fiumara}, V. and {Flaminio}, R. and {Floden}, E. and {Fong}, H. and {Font}, J.~A. and {Fornal}, B. and {Forsyth}, P.~W.~F. and {Franke}, A. and {Frasca}, S. and {Frasconi}, F. and {Frederick}, C. and {Freed}, J.~P. and {Frei}, Z. and {Freise}, A. and {Frey}, R. and {Fritschel}, P. and {Frolov}, V.~V. and {Fronz{\'e}}, G.~G. and {Fujii}, Y. and {Fujikawa}, Y. and {Fukunaga}, M. and {Fukushima}, M. and {Fulda}, P. and {Fyffe}, M. and {Gabbard}, H.~A. and {Gabella}, W.~E. and {Gadre}, B.~U. and {Gair}, J.~R. and {Gais}, J. and {Galaudage}, S. and {Gamba}, R. and {Ganapathy}, D. and {Ganguly}, A. and {Gao}, D. and {Gaonkar}, S.~G. and {Garaventa}, B. and {Garc{\'\i}a}, F. and {Garc{\'\i}a-N{\'u}{\~n}ez}, C. and {Garc{\'\i}a-Quir{\'o}s}, C. and {Garufi}, F. and {Gateley}, B. and {Gaudio}, S. and {Gayathri}, V. and {Ge}, G. -G. and {Gemme}, G. and {Gennai}, A. and {George}, J. and {George}, R.~N. and {Gerberding}, O. and {Gergely}, L. and {Gewecke}, P. and {Ghonge}, S. and {Ghosh}, Abhirup and {Ghosh}, Archisman and {Ghosh}, Shaon and {Ghosh}, Shrobana and {Giacomazzo}, B. and {Giacoppo}, L. and {Giaime}, J.~A. and {Giardina}, K.~D. and {Gibson}, D.~R. and {Gier}, C. and {Giesler}, M. and {Giri}, P. and {Gissi}, F. and {Glanzer}, J. and {Gleckl}, A.~E. and {Godwin}, P. and {Goetz}, E. and {Goetz}, R. and {Gohlke}, N. and {Golomb}, J. and {Goncharov}, B. and {Gonz{\'a}lez}, G. and {Gopakumar}, A. and {Gosselin}, M. and {Gouaty}, R. and {Gould}, D.~W. and {Grace}, B. and {Grado}, A. and {Granata}, M. and {Granata}, V. and {Grant}, A. and {Gras}, S. and {Grassia}, P. and {Gray}, C. and {Gray}, R. and {Greco}, G. and {Green}, A.~C. and {Green}, R. and {Gretarsson}, A.~M. and {Gretarsson}, E.~M. and {Griffith}, D. and {Griffiths}, W. and {Griggs}, H.~L. and {Grignani}, G. and {Grimaldi}, A. and {Grimm}, S.~J. and {Grote}, H. and {Grunewald}, S. and {Gruning}, P. and {Guerra}, D. and {Guidi}, G.~M. and {Guimaraes}, A.~R. and {Guix{\'e}}, G. and {Gulati}, H.~K. and {Guo}, H. -K. and {Guo}, Y. and {Gupta}, Anchal and {Gupta}, Anuradha and {Gupta}, P. and {Gustafson}, E.~K. and {Gustafson}, R. and {Guzman}, F. and {Ha}, S. and {Haegel}, L. and {Hagiwara}, A. and {Haino}, S. and {Halim}, O. and {Hall}, E.~D. and {Hamilton}, E.~Z. and {Hammond}, G. and {Han}, W. -B. and {Haney}, M. and {Hanks}, J. and {Hanna}, C. and {Hannam}, M.~D. and {Hannuksela}, O. and {Hansen}, H. and {Hansen}, T.~J. and {Hanson}, J. and {Harder}, T. and {Hardwick}, T. and {Haris}, K. and {Harms}, J. and {Harry}, G.~M. and {Harry}, I.~W. and {Hartwig}, D. and {Hasegawa}, K. and {Haskell}, B. and {Hasskew}, R.~K. and {Haster}, C. -J. and {Hattori}, K. and {Haughian}, K. and {Hayakawa}, H. and {Hayama}, K. and {Hayes}, F.~J. and {Healy}, J. and {Heidmann}, A. and {Heidt}, A. and {Heintze}, M.~C. and {Heinze}, J. and {Heinzel}, J. and {Heitmann}, H. and {Hellman}, F. and {Hello}, P. and {Helmling-Cornell}, A.~F. and {Hemming}, G. and {Hendry}, M. and {Heng}, I.~S. and {Hennes}, E. and {Hennig}, J. and {Hennig}, M.~H. and {Hernandez}, A.~G. and {Hernandez Vivanco}, F. and {Heurs}, M. and {Hild}, S. and {Hill}, P. and {Himemoto}, Y. and {Hines}, A.~S. and {Hiranuma}, Y. and {Hirata}, N. and {Hirose}, E. and {Hochheim}, S. and {Hofman}, D. and {Hohmann}, J.~N. and {Holcomb}, D.~G. and {Holland}, N.~A. and {Holley-Bockelmann}, K. and {Hollows}, I.~J. and {Holmes}, Z.~J. and {Holt}, K. and {Holz}, D.~E. and {Hong}, Z. and {Hopkins}, P. and {Hough}, J. and {Hourihane}, S. and {Howell}, E.~J. and {Hoy}, C.~G. and {Hoyland}, D. and {Hreibi}, A. and {Hsieh}, B. -H. and {Hsu}, Y. and {Huang}, G. -Z. and {Huang}, H. -Y. and {Huang}, P. and {Huang}, Y. -C. and {Huang}, Y. -J. and {Huang}, Y. and {H{\"u}bner}, M.~T. and {Huddart}, A.~D. and {Hughey}, B. and {Hui}, D.~C.~Y. and {Hui}, V. and {Husa}, S. and {Huttner}, S.~H. and {Huxford}, R. and {Huynh-Dinh}, T. and {Ide}, S. and {Idzkowski}, B. and {Iess}, A. and {Ikenoue}, B. and {Imam}, S. and {Inayoshi}, K. and {Ingram}, C. and {Inoue}, Y. and {Ioka}, K. and {Isi}, M. and {Isleif}, K. and {Ito}, K. and {Itoh}, Y. and {Iyer}, B.~R. and {Izumi}, K. and {Jaberianhamedan}, V. and {Jacqmin}, T. and {Jadhav}, S.~J. and {Jadhav}, S.~P. and {James}, A.~L. and {Jan}, A.~Z. and {Jani}, K. and {Janquart}, J. and {Janssens}, K. and {Janthalur}, N.~N. and {Jaranowski}, P. and {Jariwala}, D. and {Jaume}, R. and {Jenkins}, A.~C. and {Jenner}, K. and {Jeon}, C. and {Jeunon}, M. and {Jia}, W. and {Jin}, H. -B. and {Johns}, G.~R. and {Johnson-McDaniel}, N.~K. and {Jones}, A.~W. and {Jones}, D.~I. and {Jones}, J.~D. and {Jones}, P. and {Jones}, R. and {Jonker}, R.~J.~G. and {Ju}, L. and {Jung}, P. and {Jung}, K. and {Junker}, J. and {Juste}, V. and {Kaihotsu}, K. and {Kajita}, T. and {Kakizaki}, M. and {Kalaghatgi}, C.~V. and {Kalogera}, V. and {Kamai}, B. and {Kamiizumi}, M. and {Kanda}, N. and {Kandhasamy}, S. and {Kang}, G. and {Kanner}, J.~B. and {Kao}, Y. and {Kapadia}, S.~J. and {Kapasi}, D.~P. and {Karat}, S. and {Karathanasis}, C. and {Karki}, S. and {Kashyap}, R. and {Kasprzack}, M. and {Kastaun}, W. and {Katsanevas}, S. and {Katsavounidis}, E. and {Katzman}, W. and {Kaur}, T. and {Kawabe}, K. and {Kawaguchi}, K. and {Kawai}, N. and {Kawasaki}, T. and {K{\'e}f{\'e}lian}, F. and {Keitel}, D. and {Key}, J.~S. and {Khadka}, S. and {Khalili}, F.~Y. and {Khan}, S. and {Khazanov}, E.~A. and {Khetan}, N. and {Khursheed}, M. and {Kijbunchoo}, N. and {Kim}, C. and {Kim}, J.~C. and {Kim}, J. and {Kim}, K. and {Kim}, W.~S. and {Kim}, Y. -M. and {Kimball}, C. and {Kimura}, N. and {Kinley-Hanlon}, M. and {Kirchhoff}, R. and {Kissel}, J.~S. and {Kita}, N. and {Kitazawa}, H. and {Kleybolte}, L. and {Klimenko}, S. and {Knee}, A.~M. and {Knowles}, T.~D. and {Knyazev}, E. and {Koch}, P. and {Koekoek}, G. and {Kojima}, Y. and {Kokeyama}, K. and {Koley}, S. and {Kolitsidou}, P. and {Kolstein}, M. and {Komori}, K. and {Kondrashov}, V. and {Kong}, A.~K.~H. and {Kontos}, A. and {Koper}, N. and {Korobko}, M. and {Kotake}, K. and {Kovalam}, M. and {Kozak}, D.~B. and {Kozakai}, C. and {Kozu}, R. and {Kringel}, V. and {Krishnendu}, N.~V. and {Kr{\'o}lak}, A. and {Kuehn}, G. and {Kuei}, F. and {Kuijer}, P. and {Kulkarni}, S. and {Kumar}, A. and {Kumar}, P. and {Kumar}, Rahul and {Kumar}, Rakesh and {Kume}, J. and {Kuns}, K. and {Kuo}, C. and {Kuo}, H. -S. and {Kuromiya}, Y. and {Kuroyanagi}, S. and {Kusayanagi}, K. and {Kuwahara}, S. and {Kwak}, K. and {Lagabbe}, P. and {Laghi}, D. and {Lalande}, E. and {Lam}, T.~L. and {Lamberts}, A. and {Landry}, M. and {Lane}, B.~B. and {Lang}, R.~N. and {Lange}, J. and {Lantz}, B. and {La Rosa}, I. and {Lartaux-Vollard}, A. and {Lasky}, P.~D. and {Laxen}, M. and {Lazzarini}, A. and {Lazzaro}, C. and {Leaci}, P. and {Leavey}, S. and {Lecoeuche}, Y.~K. and {Lee}, H.~K. and {Lee}, H.~M. and {Lee}, H.~W. and {Lee}, J. and {Lee}, K. and {Lee}, R. and {Lehmann}, J. and {Lema{\^\i}tre}, A. and {Leonardi}, M. and {Leroy}, N. and {Letendre}, N. and {Levesque}, C. and {Levin}, Y. and {Leviton}, J.~N. and {Leyde}, K. and {Li}, A.~K.~Y. and {Li}, B. and {Li}, J. and {Li}, K.~L. and {Li}, T.~G.~F. and {Li}, X. and {Lin}, C. -Y. and {Lin}, F. -K. and {Lin}, F. -L. and {Lin}, H.~L. and {Lin}, L.~C. -C. and {Linde}, F. and {Linker}, S.~D. and {Linley}, J.~N. and {Littenberg}, T.~B. and {Liu}, G.~C. and {Liu}, J. and {Liu}, K. and {Liu}, X. and {Llamas}, F. and {Llorens-Monteagudo}, M. and {Lo}, R.~K.~L. and {Lockwood}, A. and {Loh}, M. and {London}, L.~T. and {Longo}, A. and {Lopez}, D. and {Portilla}, M. Lopez and {Lorenzini}, M. and {Loriette}, V. and {Lormand}, M. and {Losurdo}, G. and {Lott}, T.~P. and {Lough}, J.~D. and {Lousto}, C.~O. and {Lovelace}, G. and {Lucaccioni}, J.~F. and {L{\"u}ck}, H. and {Lumaca}, D. and {Lundgren}, A.~P. and {Luo}, L. -W. and {Lynam}, J.~E. and {Macas}, R. and {Macinnis}, M. and {MacLeod}, D.~M. and {MacMillan}, I.~A.~O. and {Macquet}, A. and {Hernandez}, I. Maga{\~n}a and {Magazz{\`u}}, C. and {Magee}, R.~M. and {Maggiore}, R. and {Magnozzi}, M. and {Mahesh}, S. and {Majorana}, E. and {Makarem}, C. and {Maksimovic}, I. and {Maliakal}, S. and {Malik}, A. and {Man}, N. and {Mandic}, V. and {Mangano}, V. and {Mango}, J.~L. and {Mansell}, G.~L. and {Manske}, M. and {Mantovani}, M. and {Mapelli}, M. and {Marchesoni}, F. and {Marchio}, M. and {Marion}, F. and {Mark}, Z. and {M{\'a}rka}, S. and {M{\'a}rka}, Z. and {Markakis}, C. and {Markosyan}, A.~S. and {Markowitz}, A. and {Maros}, E. and {Marquina}, A. and {Marsat}, S. and {Martelli}, F. and {Martin}, I.~W. and {Martin}, R.~M. and {Martinez}, M. and {Martinez}, V.~A. and {Martinez}, V. and {Martinovic}, K. and {Martynov}, D.~V. and {Marx}, E.~J. and {Masalehdan}, H. and {Mason}, K. and {Massera}, E. and {Masserot}, A. and {Massinger}, T.~J. and {Masso-Reid}, M. and {Mastrogiovanni}, S. and {Matas}, A. and {Mateu-Lucena}, M. and {Matichard}, F. and {Matiushechkina}, M. and {Mavalvala}, N. and {McCann}, J.~J. and {McCarthy}, R. and {McClelland}, D.~E. and {McClincy}, P.~K. and {McCormick}, S. and {McCuller}, L. and {McGhee}, G.~I. and {McGuire}, S.~C. and {McIsaac}, C. and {McIver}, J. and {McRae}, T. and {McWilliams}, S.~T. and {Meacher}, D. and {Mehmet}, M. and {Mehta}, A.~K. and {Meijer}, Q. and {Melatos}, A. and {Melchor}, D.~A. and {Mendell}, G. and {Menendez-Vazquez}, A. and {Menoni}, C.~S. and {Mercer}, R.~A. and {Mereni}, L. and {Merfeld}, K. and {Merilh}, E.~L. and {Merritt}, J.~D. and {Merzougui}, M. and {Meshkov}, S. and {Messenger}, C. and {Messick}, C. and {Meyers}, P.~M. and {Meylahn}, F. and {Mhaske}, A. and {Miani}, A. and {Miao}, H. and {Michaloliakos}, I. and {Michel}, C. and {Michimura}, Y. and {Middleton}, H. and {Milano}, L. and {Miller}, A.~L. and {Miller}, A. and {Miller}, B. and {Millhouse}, M. and {Mills}, J.~C. and {Milotti}, E. and {Minazzoli}, O. and {Minenkov}, Y. and {Mio}, N. and {Mir}, Ll. M. and {Miravet-Ten{\'e}s}, M. and {Mishra}, C. and {Mishra}, T. and {Mistry}, T. and {Mitra}, S. and {Mitrofanov}, V.~P. and {Mitselmakher}, G. and {Mittleman}, R. and {Miyakawa}, O. and {Miyamoto}, A. and {Miyazaki}, Y. and {Miyo}, K. and {Miyoki}, S. and {Mo}, Geoffrey and {Modafferi}, L.~M. and {Moguel}, E. and {Mogushi}, K. and {Mohapatra}, S.~R.~P. and {Mohite}, S.~R. and {Molina}, I. and {Molina-Ruiz}, M. and {Mondin}, M. and {Montani}, M. and {Moore}, C.~J. and {Moraru}, D. and {Morawski}, F. and {More}, A. and {Moreno}, C. and {Moreno}, G. and {Mori}, Y. and {Morisaki}, S. and {Moriwaki}, Y. and {Morr{\'a}s}, G. and {Mours}, B. and {Mow-Lowry}, C.~M. and {Mozzon}, S. and {Muciaccia}, F. and {Mukherjee}, Arunava and {Mukherjee}, D. and {Mukherjee}, Soma and {Mukherjee}, Subroto and {Mukherjee}, Suvodip and {Mukund}, N. and {Mullavey}, A. and {Munch}, J. and {Mu{\~n}iz}, E.~A. and {Murray}, P.~G. and {Musenich}, R. and {Muusse}, S. and {Nadji}, S.~L. and {Nagano}, K. and {Nagano}, S. and {Nagar}, A. and {Nakamura}, K. and {Nakano}, H. and {Nakano}, M. and {Nakashima}, R. and {Nakayama}, Y. and {Napolano}, V. and {Nardecchia}, I. and {Narikawa}, T. and {Naticchioni}, L. and {Nayak}, B. and {Nayak}, R.~K. and {Negishi}, R. and {Neil}, B.~F. and {Neilson}, J. and {Nelemans}, G. and {Nelson}, T.~J.~N. and {Nery}, M. and {Neubauer}, P. and {Neunzert}, A. and {Ng}, K.~Y. and {Ng}, S.~W.~S. and {Nguyen}, C. and {Nguyen}, P. and {Nguyen}, T. and {Quynh}, L. Nguyen and {Ni}, W. -T. and {Nichols}, S.~A. and {Nishizawa}, A. and {Nissanke}, S. and {Nitoglia}, E. and {Nocera}, F. and {Norman}, M. and {North}, C. and {Nozaki}, S. and {Siles}, J.~F. Nu{\~n}o and {Nuttall}, L.~K. and {Oberling}, J. and {O'Brien}, B.~D. and {Obuchi}, Y. and {O'Dell}, J. and {Oelker}, E. and {Ogaki}, W. and {Oganesyan}, G. and {Oh}, J.~J. and {Oh}, K. and {Oh}, S.~H. and {Ohashi}, M. and {Ohishi}, N. and {Ohkawa}, M. and {Ohme}, F. and {Ohta}, H. and {Okada}, M.~A. and {Okutani}, Y. and {Okutomi}, K. and {Olivetto}, C. and {Oohara}, K. and {Ooi}, C. and {Oram}, R. and {O'Reilly}, B. and {Ormiston}, R.~G. and {Ormsby}, N.~D. and {Ortega}, L.~F. and {O'Shaughnessy}, R. and {O'Shea}, E. and {Oshino}, S. and {Ossokine}, S. and {Osthelder}, C. and {Otabe}, S. and {Ottaway}, D.~J. and {Overmier}, H. and {Pace}, A.~E. and {Pagano}, G. and {Page}, M.~A. and {Pagliaroli}, G. and {Pai}, A. and {Pai}, S.~A. and {Palamos}, J.~R. and {Palashov}, O. and {Palomba}, C. and {Pan}, H. and {Pan}, K. and {Panda}, P.~K. and {Pang}, H. and {Pang}, P.~T.~H. and {Pankow}, C. and {Pannarale}, F. and {Pant}, B.~C. and {Panther}, F.~H. and {Paoletti}, F. and {Paoli}, A. and {Paolone}, A. and {Parisi}, A. and {Park}, H. and {Park}, J. and {Parker}, W. and {Pascucci}, D. and {Pasqualetti}, A. and {Passaquieti}, R. and {Passuello}, D. and {Patel}, M. and {Pathak}, M. and {Patricelli}, B. and {Patron}, A.~S. and {Paul}, S. and {Payne}, E. and {Pedraza}, M. and {Pegoraro}, M. and {Pele}, A. and {Arellano}, F.~E. Pe{\~n}a and {Penn}, S. and {Perego}, A. and {Pereira}, A. and {Pereira}, T. and {Perez}, C.~J. and {P{\'e}rigois}, C. and {Perkins}, C.~C. and {Perreca}, A. and {Perri{\`e}s}, S. and {Petermann}, J. and {Petterson}, D. and {Pfeiffer}, H.~P. and {Pham}, K.~A. and {Phukon}, K.~S. and {Piccinni}, O.~J. and {Pichot}, M. and {Piendibene}, M. and {Piergiovanni}, F. and {Pierini}, L. and {Pierro}, V. and {Pillant}, G. and {Pillas}, M. and {Pilo}, F. and {Pinard}, L. and {Pinto}, I.~M. and {Pinto}, M. and {Piotrzkowski}, B. and {Piotrzkowski}, K. and {Pirello}, M. and {Pitkin}, M.~D. and {Placidi}, E. and {Planas}, L. and {Plastino}, W. and {Pluchar}, C. and {Poggiani}, R. and {Polini}, E. and {Pong}, D.~Y.~T. and {Ponrathnam}, S. and {Popolizio}, P. and {Porter}, E.~K. and {Poulton}, R. and {Powell}, J. and {Pracchia}, M. and {Pradier}, T. and {Prajapati}, A.~K. and {Prasai}, K. and {Prasanna}, R. and {Pratten}, G. and {Principe}, M. and {Prodi}, G.~A. and {Prokhorov}, L. and {Prosposito}, P. and {Prudenzi}, L. and {Puecher}, A. and {Punturo}, M. and {Puosi}, F. and {Puppo}, P. and {P{\"u}rrer}, M. and {Qi}, H. and {Quetschke}, V. and {Quitzow-James}, R. and {Qutob}, N. and {Raab}, F.~J. and {Raaijmakers}, G. and {Radkins}, H. and {Radulesco}, N. and {Raffai}, P. and {Rail}, S.~X. and {Raja}, S. and {Rajan}, C. and {Ramirez}, K.~E. and {Ramirez}, T.~D. and {Ramos-Buades}, A. and {Rana}, J. and {Rapagnani}, P. and {Rapol}, U.~D. and {Ray}, A. and {Raymond}, V. and {Raza}, N. and {Razzano}, M. and {Read}, J. and {Rees}, L.~A. and {Regimbau}, T. and {Rei}, L. and {Reid}, S. and {Reid}, S.~W. and {Reitze}, D.~H. and {Relton}, P. and {Renzini}, A. and {Rettegno}, P. and {Reza}, A. and {Rezac}, M. and {Ricci}, F. and {Richards}, D. and {Richardson}, J.~W. and {Richardson}, L. and {Riemenschneider}, G. and {Riles}, K. and {Rinaldi}, S. and {Rink}, K. and {Rizzo}, M. and {Robertson}, N.~A. and {Robie}, R. and {Robinet}, F. and {Rocchi}, A. and {Rodriguez}, S. and {Rolland}, L. and {Rollins}, J.~G. and {Romanelli}, M. and {Romano}, R. and {Romel}, C.~L. and {Romero-Rodr{\'\i}guez}, A. and {Romero-Shaw}, I.~M. and {Romie}, J.~H. and {Ronchini}, S. and {Rosa}, L. and {Rose}, C.~A. and {Rosi{\'n}ska}, D. and {Ross}, M.~P. and {Rowan}, S. and {Rowlinson}, S.~J. and {Roy}, S. and {Roy}, Santosh and {Roy}, Soumen and {Rozza}, D. and {Ruggi}, P. and {Ruiz-Rocha}, K. and {Ryan}, K. and {Sachdev}, S. and {Sadecki}, T. and {Sadiq}, J. and {Sago}, N. and {Saito}, S. and {Saito}, Y. and {Sakai}, K. and {Sakai}, Y. and {Sakellariadou}, M. and {Sakuno}, Y. and {Salafia}, O.~S. and {Salconi}, L. and {Saleem}, M. and {Salemi}, F. and {Samajdar}, A. and {Sanchez}, E.~J. and {Sanchez}, J.~H. and {Sanchez}, L.~E. and {Sanchis-Gual}, N. and {Sanders}, J.~R. and {Sanuy}, A. and {Saravanan}, T.~R. and {Sarin}, N. and {Sassolas}, B. and {Satari}, H. and {Sathyaprakash}, B.~S. and {Sato}, S. and {Sato}, T. and {Sauter}, O. and {Savage}, R.~L. and {Sawada}, T. and {Sawant}, D. and {Sawant}, H.~L. and {Sayah}, S. and {Schaetzl}, D. and {Scheel}, M. and {Scheuer}, J. and {Schiworski}, M. and {Schmidt}, P. and {Schmidt}, S. and {Schnabel}, R. and {Schneewind}, M. and {Schofield}, R.~M.~S. and {Sch{\"o}nbeck}, A. and {Schulte}, B.~W. and {Schutz}, B.~F. and {Schwartz}, E. and {Scott}, J. and {Scott}, S.~M. and {Seglar-Arroyo}, M. and {Sekiguchi}, T. and {Sekiguchi}, Y. and {Sellers}, D. and {Sengupta}, A.~S. and {Sentenac}, D. and {Seo}, E.~G. and {Sequino}, V. and {Sergeev}, A. and {Setyawati}, Y. and {Shaffer}, T. and {Shahriar}, M.~S. and {Shams}, B. and {Shao}, L. and {Sharma}, A. and {Sharma}, P. and {Shawhan}, P. and {Shcheblanov}, N.~S. and {Shibagaki}, S. and {Shikauchi}, M. and {Shimizu}, R. and {Shimoda}, T. and {Shimode}, K. and {Shinkai}, H. and {Shishido}, T. and {Shoda}, A. and {Shoemaker}, D.~H. and {Shoemaker}, D.~M. and {Shyamsundar}, S. and {Sieniawska}, M. and {Sigg}, D. and {Singer}, L.~P. and {Singh}, D. and {Singh}, N. and {Singha}, A. and {Sintes}, A.~M. and {Sipala}, V. and {Skliris}, V. and {Slagmolen}, B.~J.~J. and {Slaven-Blair}, T.~J. and {Smetana}, J. and {Smith}, J.~R. and {Smith}, R.~J.~E. and {Soldateschi}, J. and {Somala}, S.~N. and {Somiya}, K. and {Son}, E.~J. and {Soni}, K. and {Soni}, S. and {Sordini}, V. and {Sorrentino}, F. and {Sorrentino}, N. and {Sotani}, H. and {Soulard}, R. and {Souradeep}, T. and {Sowell}, E. and {Spagnuolo}, V. and {Spencer}, A.~P. and {Spera}, M. and {Srinivasan}, R. and {Srivastava}, A.~K. and {Srivastava}, V. and {Staats}, K. and {Stachie}, C. and {Steer}, D.~A. and {Steinhoff}, J. and {Steinlechner}, J. and {Steinlechner}, S. and {Stevenson}, S.~P. and {Stops}, D.~J. and {Stover}, M. and {Strain}, K.~A. and {Strang}, L.~C. and {Stratta}, G. and {Strunk}, A. and {Sturani}, R. and {Stuver}, A.~L. and {Sudhagar}, S. and {Sudhir}, V. and {Sugimoto}, R. and {Suh}, H.~G. and {Sullivan}, A.~G. and {Sullivan}, J.~M. and {Summerscales}, T.~Z. and {Sun}, H. and {Sun}, L. and {Sunil}, S. and {Sur}, A. and {Suresh}, J. and {Sutton}, P.~J. and {Suzuki}, Takamasa and {Suzuki}, Toshikazu and {Swinkels}, B.~L. and {Szczepa{\'n}czyk}, M.~J. and {Szewczyk}, P. and {Tacca}, M. and {Tagoshi}, H. and {Tait}, S.~C. and {Takahashi}, H. and {Takahashi}, R. and {Takamori}, A. and {Takano}, S. and {Takeda}, H. and {Takeda}, M. and {Talbot}, C.~J. and {Talbot}, C. and {Tanaka}, H. and {Tanaka}, Kazuyuki and {Tanaka}, Kenta and {Tanaka}, Taiki and {Tanaka}, Takahiro and {Tanasijczuk}, A.~J. and {Tanioka}, S. and {Tanner}, D.~B. and {Tao}, D. and {Tao}, L. and {Mart{\'\i}n}, E.~N. Tapia San and {Taranto}, C. and {Tasson}, J.~D. and {Telada}, S. and {Tenorio}, R. and {Terhune}, J.~E. and {Terkowski}, L. and {Thirugnanasambandam}, M.~P. and {Thomas}, L. and {Thomas}, M. and {Thomas}, P. and {Thompson}, J.~E. and {Thondapu}, S.~R. and {Thorne}, K.~A. and {Thrane}, E. and {Tiwari}, Shubhanshu and {Tiwari}, Srishti and {Tiwari}, V. and {Toivonen}, A.~M. and {Toland}, K. and {Tolley}, A.~E. and {Tomaru}, T. and {Tomigami}, Y. and {Tomura}, T. and {Tonelli}, M. and {Torres-Forn{\'e}}, A. and {Torrie}, C.~I. and {E Melo}, I. Tosta and {T{\"o}yr{\"a}}, D. and {Trapananti}, A. and {Travasso}, F. and {Traylor}, G. and {Trevor}, M. and {Tringali}, M.~C. and {Tripathee}, A. and {Troiano}, L. and {Trovato}, A. and {Trozzo}, L. and {Trudeau}, R.~J. and {Tsai}, D.~S. and {Tsai}, D. and {Tsang}, K.~W. and {Tsang}, T. and {Tsao}, J. -S. and {Tse}, M. and {Tso}, R. and {Tsubono}, K. and {Tsuchida}, S. and {Tsukada}, L. and {Tsuna}, D. and {Tsutsui}, T. and {Tsuzuki}, T. and {Turbang}, K. and {Turconi}, M. and {Tuyenbayev}, D. and {Ubhi}, A.~S. and {Uchikata}, N. and {Uchiyama}, T. and {Udall}, R.~P. and {Ueda}, A. and {Uehara}, T. and {Ueno}, K. and {Ueshima}, G. and {Unnikrishnan}, C.~S. and {Uraguchi}, F. and {Urban}, A.~L. and {Ushiba}, T. and {Utina}, A. and {Vahlbruch}, H. and {Vajente}, G. and {Vajpeyi}, A. and {Valdes}, G. and {Valentini}, M. and {Valsan}, V. and {van Bakel}, N. and {van Beuzekom}, M. and {van den Brand}, J.~F.~J. and {van den Broeck}, C. and {Vander-Hyde}, D.~C. and {van der Schaaf}, L. and {van Heijningen}, J.~V. and {Vanosky}, J. and {van Putten}, M.~H.~P.~M. and {van Remortel}, N. and {Vardaro}, M. and {Vargas}, A.~F. and {Varma}, V. and {Vas{\'u}th}, M. and {Vecchio}, A. and {Vedovato}, G. and {Veitch}, J. and {Veitch}, P.~J. and {Venneberg}, J. and {Venugopalan}, G. and {Verkindt}, D. and {Verma}, P. and {Verma}, Y. and {Veske}, D. and {Vetrano}, F. and {Vicer{\'e}}, A. and {Vidyant}, S. and {Viets}, A.~D. and {Vijaykumar}, A. and {Villa-Ortega}, V. and {Vinet}, J. -Y. and {Virtuoso}, A. and {Vitale}, S. and {Vo}, T. and {Vocca}, H. and {von Reis}, E.~R.~G. and {von Wrangel}, J.~S.~A. and {Vorvick}, C. and {Vyatchanin}, S.~P. and {Wade}, L.~E. and {Wade}, M. and {Wagner}, K.~J. and {Walet}, R.~C. and {Walker}, M. and {Wallace}, G.~S. and {Wallace}, L. and {Walsh}, S. and {Wang}, J. and {Wang}, J.~Z. and {Wang}, W.~H. and {Ward}, R.~L. and {Warner}, J. and {Was}, M. and {Washimi}, T. and {Washington}, N.~Y. and {Watchi}, J. and {Weaver}, B. and {Webster}, S.~A. and {Weinert}, M. and {Weinstein}, A.~J. and {Weiss}, R. and {Weller}, C.~M. and {Weller}, R.~A. and {Wellmann}, F. and {Wen}, L. and {We{\ss}els}, P. and {Wette}, K. and {Whelan}, J.~T. and {White}, D.~D. and {Whiting}, B.~F. and {Whittle}, C. and {Wilken}, D. and {Williams}, D. and {Williams}, M.~J. and {Williams}, N. and {Williamson}, A.~R. and {Willis}, J.~L. and {Willke}, B. and {Wilson}, D.~J. and {Winkler}, W. and {Wipf}, C.~C. and {Wlodarczyk}, T. and {Woan}, G. and {Woehler}, J. and {Wofford}, J.~K. and {Wong}, I.~C.~F. and {Wu}, C. and {Wu}, D.~S. and {Wu}, H. and {Wu}, S. and {Wysocki}, D.~M. and {Xiao}, L. and {Xu}, W. -R. and {Yamada}, T. and {Yamamoto}, H. and {Yamamoto}, Kazuhiro and {Yamamoto}, Kohei and {Yamamoto}, T. and {Yamashita}, K. and {Yamazaki}, R. and {Yang}, F.~W. and {Yang}, L. and {Yang}, Y. and {Yang}, Yang and {Yang}, Z. and {Yap}, M.~J. and {Yeeles}, D.~W. and {Yelikar}, A.~B. and {Ying}, M. and {Yokogawa}, K. and {Yokoyama}, J. and {Yokozawa}, T. and {Yoo}, J. and {Yoshioka}, T. and {Yu}, Hang and {Yu}, Haocun and {Yuzurihara}, H. and {Zadro{\.z}ny}, A. and {Zanolin}, M. and {Zeidler}, S. and {Zelenova}, T. and {Zendri}, J. -P. and {Zevin}, M. and {Zhan}, M. and {Zhang}, H. and {Zhang}, J. and {Zhang}, L. and {Zhang}, T. and {Zhang}, Y. and {Zhao}, C. and {Zhao}, G. and {Zhao}, Y. and {Zhao}, Yue and {Zheng}, Y. and {Zhou}, R. and {Zhou}, Z. and {Zhu}, X.~J. and {Zhu}, Z. -H. and {Zimmerman}, A.~B. and {Zlochower}, Y. and {Zucker}, M.~E. and {Zweizig}, J. and {{LIGO Scientific Collaboration}} and {{Virgo Collaboration}} and {{KAGRA Collaboration}}},
        title = "{GWTC-3: Compact Binary Coalescences Observed by LIGO and Virgo during the Second Part of the Third Observing Run}",
      journal = {Physical Review X},
         year = 2023,
        month = oct,
       volume = {13},
       number = {4},
          eid = {041039},
        pages = {041039},
          doi = {10.1103/PhysRevX.13.041039},
archivePrefix = {arXiv},
       eprint = {2111.03606},
 primaryClass = {gr-qc},
       adsurl = {https://ui.adsabs.harvard.edu/abs/2023PhRvX..13d1039A}
}

@ARTICLE{Olejak2020,
       author = {{Olejak}, A. and {Belczynski}, K. and {Bulik}, T. and {Sobolewska}, M.},
        title = "{Synthetic catalog of black holes in the Milky Way}",
      journal = {\aap},
         year = 2020,
        month = jun,
       volume = {638},
          eid = {A94},
        pages = {A94},
          doi = {10.1051/0004-6361/201936557},
archivePrefix = {arXiv},
       eprint = {1908.08775},
 primaryClass = {astro-ph.SR},
       adsurl = {https://ui.adsabs.harvard.edu/abs/2020A&A...638A..94O}
}

@ARTICLE{Miller-Jones2021,
       author = {{Miller-Jones}, James C.~A. and {Bahramian}, Arash and {Orosz}, Jerome A. and {Mandel}, Ilya and {Gou}, Lijun and {Maccarone}, Thomas J. and {Neijssel}, Coenraad J. and {Zhao}, Xueshan and {Zi{\'o}{\l}kowski}, Janusz and {Reid}, Mark J. and {Uttley}, Phil and {Zheng}, Xueying and {Byun}, Do-Young and {Dodson}, Richard and {Grinberg}, Victoria and {Jung}, Taehyun and {Kim}, Jeong-Sook and {Marcote}, Benito and {Markoff}, Sera and {Rioja}, Mar{\'\i}a J. and {Rushton}, Anthony P. and {Russell}, David M. and {Sivakoff}, Gregory R. and {Tetarenko}, Alexandra J. and {Tudose}, Valeriu and {Wilms}, Joern},
        title = "{Cygnus X-1 contains a 21-solar mass black hole{\textemdash}Implications for massive star winds}",
      journal = {Science},
         year = 2021,
        month = mar,
       volume = {371},
       number = {6533},
        pages = {1046-1049},
          doi = {10.1126/science.abb3363},
archivePrefix = {arXiv},
       eprint = {2102.09091},
 primaryClass = {astro-ph.HE},
       adsurl = {https://ui.adsabs.harvard.edu/abs/2021Sci...371.1046M}
}

@ARTICLE{Sukhbold2016,
       author = {{Sukhbold}, Tuguldur and {Ertl}, T. and {Woosley}, S.~E. and {Brown}, Justin M. and {Janka}, H. -T.},
        title = "{Core-collapse Supernovae from 9 to 120 Solar Masses Based on Neutrino-powered Explosions}",
      journal = {\apj},
         year = 2016,
        month = apr,
       volume = {821},
       number = {1},
          eid = {38},
        pages = {38},
          doi = {10.3847/0004-637X/821/1/38},
archivePrefix = {arXiv},
       eprint = {1510.04643},
 primaryClass = {astro-ph.HE},
       adsurl = {https://ui.adsabs.harvard.edu/abs/2016ApJ...821...38S}
}

@ARTICLE{Sukhbold2014,
       author = {{Sukhbold}, Tuguldur and {Woosley}, S.~E.},
        title = "{The Compactness of Presupernova Stellar Cores}",
      journal = {\apj},
         year = 2014,
        month = mar,
       volume = {783},
       number = {1},
          eid = {10},
        pages = {10},
          doi = {10.1088/0004-637X/783/1/10},
archivePrefix = {arXiv},
       eprint = {1311.6546},
 primaryClass = {astro-ph.SR},
       adsurl = {https://ui.adsabs.harvard.edu/abs/2014ApJ...783...10S}
}

@ARTICLE{Woosley2017,
       author = {{Woosley}, S.~E.},
        title = "{Pulsational Pair-instability Supernovae}",
      journal = {\apj},
         year = 2017,
        month = feb,
       volume = {836},
       number = {2},
          eid = {244},
        pages = {244},
          doi = {10.3847/1538-4357/836/2/244},
archivePrefix = {arXiv},
       eprint = {1608.08939},
 primaryClass = {astro-ph.HE},
       adsurl = {https://ui.adsabs.harvard.edu/abs/2017ApJ...836..244W}
}

@ARTICLE{Woosley2020,
       author = {{Woosley}, S.~E. and {Sukhbold}, Tuguldur and {Janka}, H. -T.},
        title = "{The Birth Function for Black Holes and Neutron Stars in Close Binaries}",
      journal = {\apj},
         year = 2020,
        month = jun,
       volume = {896},
       number = {1},
          eid = {56},
        pages = {56},
          doi = {10.3847/1538-4357/ab8cc1},
archivePrefix = {arXiv},
       eprint = {2001.10492},
 primaryClass = {astro-ph.HE},
       adsurl = {https://ui.adsabs.harvard.edu/abs/2020ApJ...896...56W}
}

@ARTICLE{Fardeen2024,
       author = {{Fardeen}, James and {McGill}, Peter and {Perkins}, Scott E. and {Dawson}, William A. and {Abrams}, Natasha S. and {Lu}, Jessica R. and {Ho}, Ming-Feng and {Bird}, Simeon},
        title = "{Astrometric Microlensing by Primordial Black Holes with the Roman Space Telescope}",
      journal = {\apj},
         year = 2024,
        month = apr,
       volume = {965},
       number = {2},
          eid = {138},
        pages = {138},
          doi = {10.3847/1538-4357/ad3243},
archivePrefix = {arXiv},
       eprint = {2312.13249},
 primaryClass = {astro-ph.GA},
       adsurl = {https://ui.adsabs.harvard.edu/abs/2024ApJ...965..138F}
}

@ARTICLE{Kaczmarek2025,
       author = {{Kaczmarek}, Zofia and {McGill}, Peter and {Perkins}, Scott E. and {Dawson}, William A. and {Huston}, Macy and {Ho}, Ming-Feng and {Abrams}, Natasha S. and {Lu}, Jessica R.},
        title = "{On Finding Black Holes in Photometric Microlensing Surveys}",
      journal = {\apj},
         year = 2025,
        month = mar,
       volume = {981},
       number = {2},
          eid = {183},
        pages = {183},
          doi = {10.3847/1538-4357/adb1d7},
archivePrefix = {arXiv},
       eprint = {2410.14098},
 primaryClass = {astro-ph.SR},
       adsurl = {https://ui.adsabs.harvard.edu/abs/2025ApJ...981..183K}
}

@ARTICLE{Kaczmarek2026,
       author = {{Kaczmarek}, Zofia and {Halasi-Kun}, Abby and {McGill}, Peter and {Perkins}, Scott E. and {Dawson}, William A.},
        title = "{Astrometric microlensing probes of the isolated neutron star population with Roman}",
      journal = {\aap},
         year = 2026,
        month = apr,
        volume = 707,
          eid = {A264},
        pages = {A264},
          doi = {10.1051/0004-6361/202558238},
          }

@ARTICLE{Panuzzo2024,
       author = {{Gaia Collaboration} and {Panuzzo}, P. and {Mazeh}, T. and {Arenou}, F. and {Holl}, B. and {Caffau}, E. and {Jorissen}, A. and {Babusiaux}, C. and {Gavras}, P. and {Sahlmann}, J. and {Bastian}, U. and {Wyrzykowski}, {\L}. and {Eyer}, L. and {Leclerc}, N. and {Bauchet}, N. and {Bombrun}, A. and {Mowlavi}, N. and {Seabroke}, G.~M. and {Teyssier}, D. and {Balbinot}, E. and {Helmi}, A. and {Brown}, A.~G.~A. and {Vallenari}, A. and {Prusti}, T. and {de Bruijne}, J.~H.~J. and {Barbier}, A. and {Biermann}, M. and {Creevey}, O.~L. and {Ducourant}, C. and {Evans}, D.~W. and {Guerra}, R. and {Hutton}, A. and {Jordi}, C. and {Klioner}, S.~A. and {Lammers}, U. and {Lindegren}, L. and {Luri}, X. and {Mignard}, F. and {Nicolas}, C. and {Randich}, S. and {Sartoretti}, P. and {Smiljanic}, R. and {Tanga}, P. and {Walton}, N.~A. and {Aerts}, C. and {Bailer-Jones}, C.~A.~L. and {Cropper}, M. and {Drimmel}, R. and {Jansen}, F. and {Katz}, D. and {Lattanzi}, M.~G. and {Soubiran}, C. and {Th{\'e}venin}, F. and {van Leeuwen}, F. and {Andrae}, R. and {Audard}, M. and {Bakker}, J. and {Blomme}, R. and {Casta{\~n}eda}, J. and {De Angeli}, F. and {Fabricius}, C. and {Fouesneau}, M. and {Fr{\'e}mat}, Y. and {Galluccio}, L. and {Guerrier}, A. and {Heiter}, U. and {Masana}, E. and {Messineo}, R. and {Nienartowicz}, K. and {Pailler}, F. and {Riclet}, F. and {Roux}, W. and {Sordo}, R. and {Gracia-Abril}, G. and {Portell}, J. and {Altmann}, M. and {Benson}, K. and {Berthier}, J. and {Burgess}, P.~W. and {Busonero}, D. and {Busso}, G. and {Cacciari}, C. and {C{\'a}novas}, H. and {Carrasco}, J.~M. and {Carry}, B. and {Cellino}, A. and {Cheek}, N. and {Clementini}, G. and {Damerdji}, Y. and {Davidson}, M. and {de Teodoro}, P. and {Delchambre}, L. and {Dell'Oro}, A. and {Fraile Garcia}, E. and {Garabato}, D. and {Garc{\'\i}a-Lario}, P. and {Haigron}, R. and {Hambly}, N.~C. and {Harrison}, D.~L. and {Hatzidimitriou}, D. and {Hern{\'a}ndez}, J. and {Hestroffer}, D. and {Hodgkin}, S.~T. and {Jamal}, S. and {Jevardat de Fombelle}, G. and {Jordan}, S. and {Krone-Martins}, A. and {Lanzafame}, A.~C. and {L{\"o}ffler}, W. and {Lorca}, A. and {Marchal}, O. and {Marrese}, P.~M. and {Moitinho}, A. and {Muinonen}, K. and {Nu{\~n}ez Campos}, M. and {Oreshina-Slezak}, I. and {Osborne}, P. and {Pancino}, E. and {Pauwels}, T. and {Recio-Blanco}, A. and {Riello}, M. and {Rimoldini}, L. and {Robin}, A.~C. and {Roegiers}, T. and {Sarro}, L.~M. and {Schultheis}, M. and {Smith}, M. and {Sozzetti}, A. and {Utrilla}, E. and {van Leeuwen}, M. and {Weingrill}, K. and {Abbas}, U. and {{\'A}brah{\'a}m}, P. and {Abreu Aramburu}, A. and {Ahmed}, S. and {Altavilla}, G. and {{\'A}lvarez}, M.~A. and {Anders}, F. and {Anderson}, R.~I. and {Anglada Varela}, E. and {Antoja}, T. and {Baig}, S. and {Baines}, D. and {Baker}, S.~G. and {Balaguer-N{\'u}{\~n}ez}, L. and {Balog}, Z. and {Barache}, C. and {Barros}, M. and {Barstow}, M.~A. and {Bartolom{\'e}}, S. and {Bashi}, D. and {Bassilana}, J. -L. and {Baudeau}, N. and {Becciani}, U. and {Bedin}, L.~R. and {Bellas-Velidis}, I. and {Bellazzini}, M. and {Beordo}, W. and {Bernet}, M. and {Bertolotto}, C. and {Bertone}, S. and {Bianchi}, L. and {Binnenfeld}, A. and {Blanco-Cuaresma}, S. and {Bland-Hawthorn}, J. and {Blazere}, A. and {Boch}, T. and {Bossini}, D. and {Bouquillon}, S. and {Bragaglia}, A. and {Braine}, J. and {Bratsolis}, E. and {Breedt}, E. and {Bressan}, A. and {Brouillet}, N. and {Brugaletta}, E. and {Bucciarelli}, B. and {Butkevich}, A.~G. and {Buzzi}, R. and {Camut}, A. and {Cancelliere}, R. and {Cantat-Gaudin}, T. and {Capilla Guilarte}, D. and {Carballo}, R. and {Carlucci}, T. and {Carnerero}, M.~I. and {Carretero}, J. and {Carton}, S. and {Casamiquela}, L. and {Casey}, A. and {Castellani}, M. and {Castro-Ginard}, A. and {Ceraj}, L. and {Cesare}, V. and {Charlot}, P. and {Chaudet}, C. and {Chemin}, L. and {Chiavassa}, A. and {Chornay}, N. and {Chosson}, D. and {Cooper}, W.~J. and {Cornez}, T. and {Cowell}, S. and {Crosta}, M. and {Crowley}, C. and {Cruz Reyes}, M. and {Dafonte}, C. and {Dal Ponte}, M. and {David}, M. and {de Laverny}, P. and {De Luise}, F. and {De March}, R. and {de Torres}, A. and {del Peloso}, E.~F. and {Delbo}, M. and {Delgado}, A. and {Delisle}, J. -B. and {Demouchy}, C. and {Denis}, E. and {Dharmawardena}, T.~E. and {Di Giacomo}, F. and {Diener}, C. and {Distefano}, E. and {Dolding}, C. and {Dsilva}, K. and {Enke}, H. and {Fabre}, C. and {Fabrizio}, M. and {Faigler}, S. and {Fatovi{\'c}}, M. and {Fedorets}, G. and {Fern{\'a}ndez-Hern{\'a}ndez}, J. and {Fernique}, P. and {Figueras}, F. and {Fouron}, C. and {Fragkoudi}, F. and {Gai}, M. and {Galinier}, M. and {Garcia-Serrano}, A. and {Garc{\'\i}a-Torres}, M. and {Garofalo}, A. and {Gerlach}, E. and {Geyer}, R. and {Giacobbe}, P. and {Gilmore}, G. and {Girona}, S. and {Giuffrida}, G. and {Gomboc}, A. and {Gomez}, A. and {Gonz{\'a}lez-Santamar{\'\i}a}, I. and {Gosset}, E. and {Granvik}, M. and {Gregori Barrera}, V. and {Guti{\'e}rrez-S{\'a}nchez}, R. and {Haywood}, M. and {Helmer}, A. and {Hidalgo}, S.~L. and {Hilger}, T. and {Hobbs}, D. and {Hottier}, C. and {Huckle}, H.~E. and {Jim{\'e}nez-Arranz}, {\'O}. and {Juaristi Campillo}, J. and {Kaczmarek}, Z. and {Kervella}, P. and {Khanna}, S. and {Kontizas}, M. and {Kordopatis}, G. and {Korn}, A.~J. and {K{\'o}sp{\'a}l}, {\'A}. and {Kostrzewa-Rutkowska}, Z. and {Kruszy{\'n}ska}, K. and {Kun}, M. and {Lambert}, S. and {Lanza}, A.~F. and {Lebreton}, Y. and {Lebzelter}, T. and {Leccia}, S. and {Lecoutre}, G. and {Liao}, S. and {Liberato}, L. and {Licata}, E. and {Livanou}, E. and {Lobel}, A. and {L{\'o}pez-Miralles}, J. and {Loup}, C. and {Madar{\'a}sz}, M. and {Mahy}, L. and {Mann}, R.~G. and {Manteiga}, M. and {Marcellino}, C.~P. and {Marchant}, J.~M. and {Marconi}, M. and {Mar{\'\i}n Pina}, D. and {Marinoni}, S. and {Marshall}, D.~J. and {Mart{\'\i}n Lozano}, J. and {Martin Polo}, L. and {Mart{\'\i}n-Fleitas}, J.~M. and {Marton}, G. and {Mascarenhas}, D. and {Masip}, A. and {Mastrobuono-Battisti}, A. and {McMillan}, P.~J. and {Meichsner}, J. and {Merc}, J. and {Messina}, S. and {Millar}, N.~R. and {Mints}, A. and {Mohamed}, D. and {Molina}, D. and {Molinaro}, R. and {Moln{\'a}r}, L. and {Mongui{\'o}}, M. and {Montegriffo}, P. and {Monti}, L. and {Mora}, A. and {Morbidelli}, R. and {Morris}, D. and {Mudimadugula}, R. and {Muraveva}, T. and {Musella}, I. and {Nagy}, Z. and {Nardetto}, N. and {Navarrete}, C. and {Oh}, S. and {Ordenovic}, C. and {Orenstein}, O. and {Pagani}, C. and {Pagano}, I. and {Palaversa}, L. and {Palicio}, P.~A. and {Pallas-Quintela}, L. and {Pawlak}, M. and {Penttil{\"a}}, A. and {Pesciullesi}, P. and {Pinamonti}, M. and {Plachy}, E. and {Planquart}, L. and {Plum}, G. and {Poggio}, E. and {Pourbaix}, D. and {Price-Whelan}, A.~M. and {Pulone}, L. and {Rabin}, V. and {Rainer}, M. and {Raiteri}, C.~M. and {Ramos}, P. and {Ramos-Lerate}, M. and {Ratajczak}, M. and {Re Fiorentin}, P. and {Regibo}, S. and {Reyl{\'e}}, C. and {Ripepi}, V. and {Riva}, A. and {Rix}, H. -W. and {Rixon}, G. and {Robert}, G. and {Robichon}, N. and {Robin}, C. and {Romero-G{\'o}mez}, M. and {Rowell}, N. and {Ruz Mieres}, D. and {Rybicki}, K.~A. and {Sadowski}, G. and {Sagrist{\`a} Sell{\'e}s}, A. and {Sanna}, N. and {Santove{\~n}a}, R. and {Sarasso}, M. and {Sarmiento}, M.~H. and {Sarrate Riera}, C. and {Sciacca}, E. and {S{\'e}gransan}, D. and {Semczuk}, M. and {Shahaf}, S. and {Siebert}, A. and {Slezak}, E. and {Smart}, R.~L. and {Snaith}, O.~N. and {Solano}, E. and {Solitro}, F. and {Souami}, D. and {Souchay}, J. and {Spitoni}, E. and {Spoto}, F. and {Squillante}, L.~A. and {Steele}, I.~A. and {Steidelm{\"u}ller}, H. and {Surdej}, J. and {Szabados}, L. and {Taris}, F. and {Taylor}, M.~B. and {Teixeira}, R. and {Tepper-Garcia}, T. and {Thuillot}, W. and {Tolomei}, L. and {Tonello}, N. and {Torra}, F. and {Torralba Elipe}, G. and {Trabucchi}, M. and {Trentin}, E. and {Tsantaki}, M. and {Turon}, C. and {Ulla}, A. and {Unger}, N. and {Valtchanov}, I. and {Vanel}, O. and {Vecchiato}, A. and {Vicente}, D. and {Villar}, E. and {Weiler}, M. and {Zhao}, H. and {Zorec}, J. and {Zucker}, S. and {{\v{Z}}upi{\'c}}, A. and {Zwitter}, T.},
        title = "{Discovery of a dormant 33 solar-mass black hole in pre-release Gaia astrometry}",
      journal = {\aap},
         year = 2024,
        month = jun,
       volume = {686},
          eid = {L2},
        pages = {L2},
          doi = {10.1051/0004-6361/202449763},
archivePrefix = {arXiv},
       eprint = {2404.10486},
 primaryClass = {astro-ph.GA},
       adsurl = {https://ui.adsabs.harvard.edu/abs/2024A&A...686L...2G}
}

@ARTICLE{El-Badry2023a,
       author = {{El-Badry}, Kareem and {Rix}, Hans-Walter and {Quataert}, Eliot and {Howard}, Andrew W. and {Isaacson}, Howard and {Fuller}, Jim and {Hawkins}, Keith and {Breivik}, Katelyn and {Wong}, Kaze W.~K. and {Rodriguez}, Antonio C. and {Conroy}, Charlie and {Shahaf}, Sahar and {Mazeh}, Tsevi and {Arenou}, Fr{\'e}d{\'e}ric and {Burdge}, Kevin B. and {Bashi}, Dolev and {Faigler}, Simchon and {Weisz}, Daniel R. and {Seeburger}, Rhys and {Almada Monter}, Silvia and {Wojno}, Jennifer},
        title = "{A Sun-like star orbiting a black hole}",
      journal = {\mnras},
         year = 2023,
        month = jan,
       volume = {518},
       number = {1},
        pages = {1057-1085},
          doi = {10.1093/mnras/stac3140},
archivePrefix = {arXiv},
       eprint = {2209.06833},
 primaryClass = {astro-ph.SR},
       adsurl = {https://ui.adsabs.harvard.edu/abs/2023MNRAS.518.1057E}
}

@ARTICLE{Remillard2006,
       author = {{Remillard}, Ronald A. and {McClintock}, Jeffrey E.},
        title = "{X-Ray Properties of Black-Hole Binaries}",
      journal = {\araa},
         year = 2006,
        month = sep,
       volume = {44},
       number = {1},
        pages = {49-92},
          doi = {10.1146/annurev.astro.44.051905.092532},
archivePrefix = {arXiv},
       eprint = {astro-ph/0606352},
 primaryClass = {astro-ph},
       adsurl = {https://ui.adsabs.harvard.edu/abs/2006ARA&A..44...49R}
}

@ARTICLE{Dominik2000,
       author = {{Dominik}, Martin and {Sahu}, Kailash C.},
        title = "{Astrometric Microlensing of Stars}",
      journal = {\apj},
         year = 2000,
        month = may,
       volume = {534},
       number = {1},
        pages = {213-226},
          doi = {10.1086/308716},
       adsurl = {https://ui.adsabs.harvard.edu/abs/2000ApJ...534..213D}
}
\bibliographystyle{aa}

\end{document}